\documentclass {elsart}
\usepackage{graphicx}

\usepackage{graphics}
\usepackage{graphicx}
\usepackage{epsfig}

\usepackage{amssymb}
\usepackage{amsmath}
\usepackage{float}
\usepackage{caption}
\usepackage{float}
\usepackage{graphics}
\usepackage{subfigure,epsfig}
\begin{document}

\begin{frontmatter}



\title{Fisher Information Framework for Time Series Modeling}


\author[SRC]{R. C. Venkatesan\corauthref{cor}}
\corauth[cor]{Corresponding author.}
\ead{ravi@systemsresearchcorp.com;ravicv@eth.net}
\author[UNLP]{A. Plastino}
\ead{plastino@fisica.unlp.edu.ar}

\address[SRC]{Systems Research Corporation,
Aundh, Pune 411007, India}
\address[UNLP]{IFLP, National University La Plata \&
National Research   (CONICET)\\ C. C., 727 1900, La Plata,
Argentina}

\begin{abstract}
A  robust prediction model invoking the Takens embedding theorem, whose \textit{working hypothesis} is obtained via an inference procedure based on the minimum Fisher information principle, is presented.   The coefficients of the ansatz, central to the \textit{working hypothesis} satisfy a time independent Schr\"{o}dinger-like equation in a vector setting.   The inference of i) the  probability density function of the coefficients of the \textit{working hypothesis} and ii) the establishing of constraint driven pseudo-inverse condition for the modeling phase of the prediction scheme, is made, for the case of normal distributions, with the aid of the quantum mechanical virial theorem. The well-known reciprocity relations and the associated Legendre transform structure for the Fisher information measure (FIM, hereafter)-based model in a vector setting (with least square constraints) are self-consistently derived.  These relations are demonstrated to yield an intriguing form of the FIM for the modeling phase, which defines the \textit{working hypothesis}, solely in terms of the observed data.   Cases for prediction employing time series' obtained from the: $(i)$ the Mackey-Glass delay-differential equation, $(ii)$ one ECG sample from the MIT-Beth Israel Deaconess Hospital (MIT-BIH) cardiac arrhythmia database, and $(iii)$ one ECG from the Creighton University ventricular tachyarrhythmia database.  The ECG samples were obtained from the Physionet online repository. These examples demonstrate the efficiency of  the prediction model.  Numerical examples for exemplary cases are provided.
\end{abstract}

\begin{keyword}
Fisher information \sep time series prediction \sep working hypothesis inference \sep minimum Fisher information \sep Takens theorem \sep generalized vector Fisher-Euler theorem \sep Legendre transform structure \sep Mackey-Glass equation \sep ECG's.

PACS: 05.20.-y; \ 2.50.Tt; \ 0.3.65.-w; \ 05.45.Tp
\end{keyword}

\end{frontmatter}

\section{Introduction}
Devising methods for analyzing and predicting time series is currently considered one of the most important challenges in chaotic time-series analysis (eg., see
Refs. [1-3]). In general, chaotic behavior is observed in relation with nonlinear differential equations and maps on manifolds. \textit{Times series may be construed as being the projections of manifolds onto coordinate axes}.  Much work in nonlinear dynamics has focused on the building of appropriate model(s) of the underlying physical process from a time series, with the objective of predicting the \textit{near-future} behavior of dynamical systems. The first step in formulating predictive models is that of  specifying/estimating a suitably parameterized nonlinear function of the observation.  This is followed by estimating the parameters of this function.

In general, prediction models are formulated on the basis of the
systematic and accurate identification of a \textit{working
hypothesis} [4]. This hypothesis is represented by a set of
parameters that form an ansatz.  This paper obtains the
coefficients of such an ansatz, which possess
information about the data set(s), via recourse to a Fisher information measure (FIM, hereafter) based inference procedure.

The leitmotif for obtaining the \textit{working hypothesis} by employing an inference
procedure is to formulate a  prediction model, based on  the
famed embedding theorem of Takens [5, 6].  The conceptual sophistication underlying the Takens' theorem renders the prediction problem to become an instance of extrapolation.  Currently, some of the prominent prediction models based on information theory (IT, hereafter) are: $(i)$ the
framework of Plastino et. al. [7-10] using the maximum entropy (MaxEnt, hereafter) method of Jaynes [11], and $(ii)$ the nonparametric models by Principe et. al, (eg.
see [12, 13]).

The work presented herein belongs to a class of models known as \textit{pseudo-inverse} models, for reasons described in Section 2 and 3 of this paper.  Such models have been successfully employed to forecasting tasks in a number of disciplines which include nonlinear dynamical systems [7, 8], financial data forecasting [8], prediction of tonic-clonic epileptic seizures from real-time electroencephalogram (EEG) data [9], and even fraud analysis (the London Interbank Offered Rate (LIBOR) manipulations) [10].

Generally, predictive models are of two types, viz. \textit{global} and \textit{local} (see for example Ref. [14]). Global models are based on training data collected from across the phase space.  On the other hand, in local models, the training is accomplished by measurements providing data lying in the immediate vicinity of a specific/localized region of the phase space.  \textit{Pseudo-inverse predictive models, including the one presented herein, are essentially global models which possess local characteristics} [10, 15].  Time series prediction has its roots in the theory of optimal filtering by Wiener [16]. In recent times, forecasting of chaotic time series has hitherto largely utilized artificial neural networks (ANN's, hereafter) and other learning paradigms.  Commencing from the seminal radial basis function model of Casdagli [17], some of the notable attempts to study chaotic time series comprise (but are not limited to) the time delayed neural network architectures [18], recurrent ANN's [19], maximum entropy ANN's [20], and support vector machines [21].  Within the perspective of physics-based models, the works of Crutchfield and McNamara [22] and Farmer and Sidorowich [23] constitute some of the most prominent efforts.

FIM-based studies have recently been acquiring prominence across a spectrum of disciplines ranging from physics and biology to economics (for eg., see [24]).  The prediction model presented in this paper comprises of two phases: $(i)$ the modeling phase and $(ii)$ the prediction phase.  The task of the modeling phase is to obtain the coefficients of the ansatz that suitably parameterizes
the nonlinear function of the observed time series (see Section 2 of this paper).  This phase establishes the \textit{working hypothesis}, and is accomplished with the assistance of the training data.  The prediction phase then generates forecasts based on the set of coefficients obtained in the modeling phase.

The leitmotif for the FIM-based model employed in this paper is two-fold.  First, it provides the framework to endow the modeling phase with a quantum mechanical (QM, hereafter) connotation.  This is in accordance with Wheeler's hypothesis of establishing
an information-theoretical foundation for the fundamental theories of physics [25], and is accomplished by recourse to the minimum Fisher information (MFI, hereafter) principle of H\"{u}ber [26, 27].  Variational extremization of the FIM subject to least squares constraints results in a Sturm-Liouville equation in a vector setting, hereinafter referred to as the time independent Schr\"{o}dinger-like equation.  Consequently, i) the probability density function (pdf, hereafter) of the coefficients of the ansatz, and ii) the constraint driven pseudo-inverse condition (that yields the inferred estimate coefficients, fundamental for the \textit{working hypothesis}), can be specified not only via  Gaussian (Maxwell-Boltzmann) pdf's [which are \textit{equilibrium} distributions], but also in terms of \textit{non-equilibrium} distributions [24, 28-30],
 comprising of Hermite-Gauss polynomials.\footnote{In this paper the terms Gaussian pdf and normal pdf are employed interchangeably}  This greatly widens the scope of the works presented in Refs. [7-10], and is accomplished in this paper with the aid of the QM virial theorem [31, 32] for normal distributions. Note that in inference problems involving the FIM, the Gaussian pdf's are obtained as solutions to the lowest eigenvalue by solving the time-independent  Schr\"{o}dinger-like equation  in Section 3 of this paper as an eigenvalue problem, and correspond to the \textit{ground state} of the physical Schr\"{o}dinger  wave equation (SWE, hereafter).  Further, the non-equilibrium pdf's correspond to the higher-order eigenvalue solutions of such SWE, and are linked to  \textit{excited states} of  the physical SWE (see, for eg. [33, 34]).  From a \textit{practical} perspective, this enables the performance of the modeling phase and the concomitant prediction phase to be systematically categorized in terms of an established physics-based framework.

 Next, the reciprocity relations and the Legendre transform structure (LTS, hereafter), together with the concomitant information theoretic relations for the FIM, in a vector setting and for least squares constraints, are derived.  Prior studies have derived reciprocity relations and LTS for the FIM model [35] and have analyzed such
 relations [36-39].  Recently, these works have been qualitatively extended to the case of the relative Fisher information (RFI, hereafter) [40-42] by Venkatesan and Plastino by deriving the reciprocity relations and LTS [43].  A connection between the celebrated Hellmann-Feynman theorem,  the reciprocity relations, and LTS for the RFI has been established in [44], in addition with a unique inference procedure to obtain the energy eigenvalue without recourse to solving the time-independent Schr\"{o}dinger-like equation.  These prior works differ from the analysis presented in this paper in two significant aspects - $(i)$ they treat the scalar case and $(ii)$ the prior knowledge encoded in the observed data are introduced as constraints into the variational extremization procedure in the form of expectations of the powers of the scalar independent variable.

The reciprocity relations and the LTS for the time-independent Schr\"{o}dinger-like equation derived in this paper, despite possessing a vector form and least squares constraints, mathematically resemble those derived in [35]. This augurs well with regards to the possibility of translating the entire mathematical structure of thermodynamics into the Fisher-based model presented in this paper.  The distinctions in the reciprocity relations and LTS derived in this paper vis-\'{a}-vis earlier referenced works [36-39] result in the information theoretic relations derived from these relations being qualitatively different from those obtained in the scalar case.  This fact evidences  the distinction between the results presented in this paper
and those demonstrated in Refs. [36-39] , based on  physics and on systems' theoretic [45] considerations.  \textit{Of interest is an expression that infers the FIM of the modeling phase just in terms   the observed data, hereafter referred to as the empirical FIM}.  Such relation, which is a solution of a linear PDE derived from the reciprocity relations together with the LTS that infers the FIM without recourse to the time-independent Schr\"{o}dinger-like equation, has no equivalent in the MaxEnt model.

The goals of this paper are \begin{itemize}

\item $(i)$ to provide an overview of the solution procedure.  This is done in Section 2,
\item$(ii)$  to:  $(a)$  introduce the MFI principle in a vector setting and using least square constraints, $(b)$  derive a systematic procedure for the inference of exponential pdf's of the modeling phase with the aid of the QM virial theorem, and $(c)$  obtain the constraint driven pseudo-inverse condition that yields the estimate of the coefficients comprising the \textit{working hypothesis} (see Section 2 of this paper) by invoking the QM virial theorem.  This three-fold objective is performed in Section 3.  Note that for normal pdf's the solutions of the MFI and MaxEnt principles are known to coincide [24, 46].  This paper focuses on the normal distribution to demonstrate that the results of the MaxEnt model can be derived from QM considerations and interpreted within the framework of estimation theory, which is not possible within the ambit of the MaxEnt framework,
\item $(iii)$ to derive the reciprocity relations and the LTS for the FIM in a vector setting using square constraints, analyzing the concomitant information theoretic relations.  The \textit{empircal FIM} is derived, and a preliminary analysis of its properties is performed.  This is accomplished in Section 4,
\item $(iv)$  to computationally demonstrate the efficacy of the prediction framework for the Mackey-Glass (M-G, hereafter) delay differential equation (DDE, hereafter) [47], for a 5 minute electrocardiogram (ECG, hereafter) segment of Record 207 of the MIT-Beth Israel Deaconess Hospital (MIT-BIH, hereafter) arrythmia database [48] (considered to be one of the most challenging Records in the MIT-BIH arrhythmia database) for the Modified Lead II (MLII, hereafter), and for the single ECG signal in Record cudb/cu02 of the around 8.5 minute Creighton University ventricular tachyarrhythmia (VTA, hereafter) database [49].  The ECG data are obtained from the Physionet online repository [50].  This is demonstrated in Section 5 of this paper.  The leitmotif of this exercise is as follows.  \\
    \\
    An obvious practical advantage of the pseudo-inverse model presented in this paper over a least squares approach in ordinary Euclidian space is that the former  requires the Moore-Penrose pseudo-inverse [51] of the embedding matrix $\textbf{W}$ (defined in Section 2 of this paper) and therefore, the estimate of the coefficients of the ansatz comprising the $\textit{working hypothesis}$ $<\textbf{a}>$ (see Sections 2 and 3 of this paper) derived via inference from the training data. This can be achieved even when \textbf{W} is nearly singular. The fact that the estimates $<\textbf{a}>$  are defined even when \textbf{W} is singular (or nearly singular) can in principle result in very volatile forecasts, on account of ill-conditioning.  Note that ill-conditioning \text{could} occur in the presence of a near-singular \textbf{W}, which in turn might occur if many lags of the observed data $\textbf{v}$ are present.\\
\\		
The leitmotif for the choice of the benchmarks on which to test the prediction model is as follows.  The M-G equation with delay $\tau > 14 secs.$ has a high embedding dimension [18].  Thus \textbf{W} displays more lags as compared to most prominent models describing low dimensional chaos [3, 14].  As is described in Section 2 of this paper, the rationale being that the number of lags in \textbf{W} depends upon the embedding dimension.  As  evidenced in Section 5 of this paper, the forecast of the M-G DDE is stable and accurate.  Next, ECG’s of patients suffering from serious cardiac related ailments possess artifacts which are representative of various conditions of a diseased heart.  These artifacts are noted in the reference annotations as episodes (transients).  It is demonstrated that even for the most challenging cases, the model presented in this paper accurately forecasts these episodes without any signs of volatility, thereby demonstrating the accuracy and robustness of the pseudo-inverse model.  This is established for cases where the original signal possesses highly erratic/volatile behavior.
  \end{itemize}
Numerical examples for exemplary
cases are provided. To the best of the authors' knowledge, these objectives have never hitherto been accomplished.
\section{Overview of the solution procedure}
\subsection{Basics of embedding theory}
 Given a signal \textbf{x} from an unknown dynamical system $D :
\Re^S \rightarrow \Re^S$ , the corresponding time series consists
of a sequence of $N$ \textit{stroboscopic} measurements: $ \left\{
{v\left( {\tau {}_0} \right),v\left( {\tau {}_0 + \tau _s }
\right),...,v\left( {\tau {}_0 + N\tau _s } \right)} \right\}$
made at intervals $\tau_s$.  The state space is reconstructed
using the time delay embedding [1, 5, 6], which uses a collection
of coordinates with time lag to create a vector in $d$-dimensions,
on a system considered to be in a state described by
$\textbf{x}(t) \in \Re^S$ at discrete times
\begin{equation}
\textbf{v}\left( t_n \right) = \left\{ {v\left( t_n \right),v\left( {t_n - \Delta } \right),...,v\left( {t_n - \left( {d - 1} \right)\Delta } \right)} \right\}
\end{equation}
where $\Delta=\tau_s$ is the time lag, and $d$ is the embedding dimension of the reconstruction.
It is known from Takens' theorem (eg. see Refs. [5, 6]) that for flows evolving to compact attracting manifolds of dimension $d_a$; if $d>2d_a$ for the forecasting time $T \in \Re$, $T > 0$ (time samples in this paper), there exists a functional form
of the type
\begin{equation}
\textbf{v}(t+T)=\Im (\textbf{v}(t)).\\
\end{equation}
where
\begin{equation}
\textbf{v}(t) = [v_1(t), v_2(t),..., v_d(t)],
\end{equation}
and $v_{i}(t)=v\left ( t-\left ( i-1 \right  )\Delta \right );i=1,...,d$.  A \textit{non-unique} ansatz for the mapping function of this form (employing the Einstein summation convention) is specified as [9]
\begin{equation}
\begin{array}{l}
\Im ^{\ast }\left ( \textbf{v}\left ( t \right ) \right )=a_{0}+a_{i_1}v_{i_1}+a_{i_1i_2}v_{i_1}v_{i_2}+a_{i_1i_2i_3}v_{i_1}v_{i_2}v_{i_3}+...+a_{i_1i_2i_3...i_{np}}v_{i_1}v_{i_2}v_{i_3}...v_{i_{np}},
\end{array}
\end{equation}
where $1 \le i_k \le d$ and $np$ is the polynomial degree chosen to expand the mapping $\Im^*$. The number of parameters in (4)
corresponding to $k$ terms (the degree), is the combination with repetitions
\begin{equation}
\binom{d}{k}^*=\frac{(d+k-1)!}{k!(d-1)!}.
\end{equation}
The length of the vector of parameters, $\textbf{a}$ is
\begin{equation}
N_{c}=\sum_{k=1}^{np}\binom{d}{k}^*.
\end{equation}
Other forms of ansatz' are encountered in [52].  It is important to note that specifying an ansatz of a form, such as that defined in (4), has its roots in signal processing [53].
\subsection{The modeling phase}
As an information recovery criterion, the vector of coefficients $\textbf{a}$ is obtained via inference by invoking the MFI principle. The objective is to achieve a model possessing high
predictive ability. Computations are made on the basis of the information
given by $M$ points of the time series.  These constitute the \textit{training data} obtained from the observed signal, whose utility is to infer the coefficients $\textbf{a}$.
\begin{equation}
\left [ \textbf{v}(t_n),v(t_n+T) \right ]; n=1,...,M.
\end{equation}
Given the data set (7), the parametric mapping (2) can be re-stated as
\begin{equation}
v(t_n+T)=\Im^*(\textbf{v}(t_n)); n=1,...,M.
\end{equation}
Here, (7) can be expressed in vector-matrix form as
\begin{equation}
\textbf{W}\textbf{a}=\textbf{v}_T,
\end{equation}
where $(\textbf{v}_T)_n=v(t_n+T)$ and $\textbf{W}$ is a rectangular matrix with dimensions $M \times N_c$, and whose $n^{th}$ row is:$\left [ 1,v_{i_1}(t_n),v_{i_2}(t_n)v_{i_2}(t_n),...,v_{i_2}(t_n)v_{i_2}(t_n)...v_{i_{np}}(t_n) \right ]$.  The \textit{working hypothesis} is established in Section 3 via inference of the coefficients from the observed data by invoking the MFI principle.  Here, $\textbf{a}=\left[a_{0},a_{i_1},a_{i_1i_2},..,a_{i_1i_2i_3...i_{np}} \right ]$.   It is assumed that the probability associated with $\textbf{a}$ is $f(\textbf{a})$.  Note that $\textbf{a}$ is assumed to be a continuous
random variable.  Alternately, $f(\textbf{a})$  may be defined as the \textit{empirical distributions} of the observations $\textbf{v}(t_n); n=1,...,M$ [54].   The FIM is extremized subject to the constraints
\begin{equation}
\textbf{W}<\textbf{a}>=\textbf{v}_T,
\end{equation}
and the normalization condition
\begin{equation}
\int f(\textbf{a})d\textbf{a}=1.
\end{equation}
Note that $d\textbf{a} = da_1da_2...da_{N_C}$, where $N_C$ is the number of parameters of the model. Also $<\bullet>$ denotes the expectation evaluated with respect to $f(\mathbf{\textbf{a}})$.  Section 3 derives the constraint driven pseudo-inverse condition for normal distributions by invoking the QM virial theorem as
\begin{equation}
<\textbf{a}>=\textbf{W}^\dagger \textbf{v}_T,
\end{equation}
where: $\textbf{W}^\dagger = \textbf{W}^T(\textbf{W}\textbf{W}^T)^{-1}$ is the Moore-Penrose \textit{pseudo-inverse} [51]. Note that as stated in Sections 1 and 3, unlike the MaxEnt model the FIM-based framework presented herein also allows for $f(\textbf{a})$ described by Hermite-Gauss solutions.  Such extensions of the present model and the subsequent effects on the pseudo-inverse condition are beyond the scope of this paper, and will be presented elsewhere.
\subsection{The prediction phase}
The \textit{prediction phase} commences once the pertinent parameters $<\textbf{a}>$ are determined from the $M$ training data in the \textit{modeling phase}.  These are employed to predict \textit{new}
series values
\begin{equation}
\hat{\textbf{v}}(t_n+T)_{n=1,...,M_P}=\hat{\textbf{W}}<\textbf{a}>,
\end{equation}
where $\hat{\textbf{W}}$ is a matrix of dimension $M_P \times N_C$. Note that $M_P$ is such that $M_P-M$ \textit{new} time series values may be evaluated after the training data has been reconstructed.  The prediction phase is essentially the implementation of (10), for temporal indices $n=1,...,M_P$, where, $M_P>>M$ is the sum of both the training data and the \textit {new} data to be predicted \textit{after} completion of the modeling phase.

\textit{It is important to note that the process of inference necessitates the re-definition of the \textit{working hypothesis} to account for (10) now superseding (9).  The obvious reason being that the process of inference can only evaluate $<\textbf{a}>$ and not $\textbf{a}$}.  The value of $M_P$ should  be suitably bounded to facilitate the comparison between the predicted signal obtained from the solution of (13), with the original signal.  This is done in order to judge the fidelity of the prediction through \textit{both} visual inspection and analysis; viz. calculation of the mean  squared error (MSE, hereafter) between the original and the predicted signal.  In this paper, given the original signal represented by the column vector $\textbf{Z}$, $
M_P  = dimension\left[ {\textbf{Z}} \right] - \max \left\{ {T,d} \right\}$.  \textit{Note that this non-unique fiduciary bound does not in any way constrain the IT-based prediction model, and there is nothing that prevents the value of $M_P$ from exceeding this bound should the situation require it}.  To evaluate the MSE, defining the exact measurement from the original signal as $\textbf{v}$, and the corresponding results of the predictive model as $\hat {\textbf{v}}$
\begin{equation}
MSE = \frac{1}{{M_P }}\sum\limits_{j = 1}^{M_P } {\left( {v_j  - \hat{v}_j  } \right)^2 }; j=1,...,M_P.
\end{equation}
\section{Inference framework for the modeling phase}
\subsection{The MFI principle}
Consider the probability
\begin{equation}
f\left( {\textbf{a};\theta } \right),
\end{equation}
where $\theta$ is a vector parameter.  Specializing the focus to a class of probability distributions exhibiting translational invariance where $
f(\textbf{a};\theta)=f\left( {\textbf{a} - \theta } \right)$, and assuming without any loss of generality, that the elements of the vector \textbf{a} are \textit{a priori iid}, the FI matrix with vector entries acquires the form of a diagonal matrix.  The FIM [24, 55] now takes the form
\begin{equation}
\begin{array}{l}
\textbf{I}\left[ f \right] = \int {\frac{1}{{f\left( \textbf{a} \right)}}} \left( {\frac{{d f\left( \textbf{a} \right)}}{{d \textbf{a}}}} \right)^2d\textbf{a}=\sum\limits_i {\int {f\left( \textbf{a} \right)} \left( {\frac{{\partial \ln f\left( \textbf{a} \right)}}{{\partial a_i }}} \right)^2 d\textbf{a}}= \sum\limits_i {\int {\frac{1}{{f_i \left( {a_i } \right)}}\left( {\frac{{\partial f_i }}{{\partial a_i }}} \right)^2 da_i } }=\sum\limits_i {I_i }.
 \end{array}
\end{equation}
Note that in (16), for iid entries of the vector \textbf{a}, the FIM is the trace of the FI matrix which is identical to the scalar case [24], and $I_i$ is the $ii^{th}$ diagonal element of the diagonal FI matrix.  The derivation of (16) is described in the Appendix.
With the aid of real valued amplitudes defined by
\begin{equation}
f\left( \textbf{a} \right) = \psi ^2 \left( \textbf{a} \right),
\end{equation}
the  FIM (16) may be compactly expressed as
\begin{equation}
\textbf{I}\left[ \psi \right] = 4\int {\left( {\frac{{d \psi \left( \textbf{a} \right)}}{{d \textbf{a}}}} \right)^2 d\textbf{a}}
\end{equation}
which is extremized subject to the constraint defined by Eq. (10) and the normalization condition Eq. (11), in Section 2.

A Lagrangian can be specified of the form
\begin{equation}
J\left[ \psi \right] = \int {\left\{ {4\left( {\frac{{d\psi \left( \textbf{a} \right)}}{{d\textbf{a}}}} \right)^2  - \mathop \lambda \limits^ \to  \textbf{W}\textbf{a}\psi ^2 \left( \textbf{a} \right) - \lambda _0 \psi ^2 \left( \textbf{a} \right)} \right\}d\textbf{a}},
\end{equation}
where $\vec{\lambda}$ is the vector of Lagrange multipliers associated with the constraint (10).   Eq. (19) is re-expressed with its constraint  terms described in component-wise form as
\begin{equation}
J\left[ \psi \right] = \int {\left\{ {4\left( {\frac{{d\psi \left( \textbf{a} \right)}}{{d\textbf{a}}}} \right)^2  - \sum\limits_{k = 1}^M {\lambda _k \sum\limits_{i = 1}^{N_c } {W_{ki} } } a_i \psi ^2 \left( \textbf{a} \right) - \lambda _0 \psi ^2 \left( \textbf{a} \right)} \right\}d\textbf{a}}
\end{equation}
Variational
extremization of (19) with respect to $\psi(\mathbf{a})$, and multiplying the resultant by 2 yields
\begin{equation}
 - \frac{{d^2 \psi \left( \textbf{a} \right)}}{{d\textbf{a}^2 }}\underbrace { - \frac{{\vec \lambda }}{4}\textbf{W}\textbf{a}\psi \left( \textbf{a} \right)}_{U\left( \textbf{a} \right)} = \frac{{\lambda _0 }}{4}\psi \left( \textbf{a} \right),
\end{equation}
where $U\left( \textbf{a} \right)=
 - \frac{{\vec \lambda }}{4}Wa\psi \left( a \right)$ is the empirical pseudo-potential.  Here, (21) bears a resemblance to the SWE in a vector setting with $
\frac{{\hbar ^2 }}{{2m}} = 1$.
\subsection{Inference of normal distributions and derivation of the pseudo-inverse condition}
Redefining (18) in terms of the
 pdf $f(\textbf{a})=\psi^2(\textbf{a})$ one finds after invoking the QM virial theorem [31]
\begin{equation}
\int {f\left( a \right)\left( {\frac{{d\ln f\left( \textbf{a} \right)}}{{da}}} \right)^2 d\textbf{a}}  = 4\int {f\left( \textbf{a} \right)\left( {\textbf{a}\frac{{d  U(\textbf{a})
}}{{da}}} \right)} da.
\end{equation}
Eqs. (22) yields
\begin{equation}
\int {f\left( \textbf{a} \right)} \left[ {\left( {\frac{{d\ln f\left( \textbf{a} \right)}}{{d\textbf{a}}}} \right)^2  - 4\textbf{a}\frac{{dU\left( \textbf{a} \right)}}{{d\textbf{a}}}} \right]d\textbf{a} = 0.
\end{equation}
Substituting the expression for $U(\textbf{a})$ in (21) into (23) results in
\begin{equation}
\int {f\left( \textbf{a} \right)} \left[ {\left( {\frac{{d\ln f\left( \textbf{a} \right)}}{{d\textbf{a}}}} \right)^2  + \mathop \lambda \limits^ \to  W\textbf{a}} \right]d\textbf{a} = 0.
\end{equation}
Solving (24) yields
\begin{equation}
f\left( \textbf{a} \right) = \exp \left[ { \mp \int {\sqrt { - \mathop \lambda \limits^ \to  \textbf{W}\textbf{a}} } d\textbf{a}} \right].
\end{equation}
Setting
\begin{equation}
\mathop \lambda \limits^ \to   =  - \frac{{\textbf{W}\textbf{a}}}{{\sigma ^4 }}
\end{equation}
results in the pdf
\begin{equation}
\begin{array}{l}
f\left( \textbf{a} \right) = \exp \left[ { - \int {\sqrt {\left( {\frac{{\textbf{Wa}}}{{\sigma ^4 }}} \right)^2 } } d\textbf{a}} \right]
 =
\frac{{\exp \left[ { - \frac{{\textbf{W}\textbf{a}^T \textbf{a}}}{{2\sigma ^2 }}} \right]}}{{\tilde Z}} = \frac{{\exp \left[ { - \frac{{\sum\limits_{k = 1}^M {W_{ki} \sum\limits_{i = 1}^{N_C } {a_i^2 } } }}{{2\sigma ^2 }}} \right]}}{{\tilde Z}}
; \\
\\
\tilde Z =
\int {\exp \left[ { - \frac{{\textbf{W}\textbf{a}^T \textbf{a}}}{{2\sigma ^2 }}} \right]d\textbf{a}} .
\end{array}
\end{equation}
Note that $\sigma^2$ denotes the \textit{statistical dispersion}, and $\tilde Z$ is the canonical partition function.  The above analysis is presented in a more familiar form by invoking the translational invariance property of the FIM by specifying
\begin{equation}
\textbf{r} = \textbf{a} - \left\langle \textbf{a} \right\rangle.
\end{equation}
Here, (28) has the effect of transforming (21) to
\begin{equation}
 - \frac{{d^2 \tilde \psi \left( \textbf{r} \right)}}{{d\textbf{r}^2 }} + \tilde U\left( \textbf{r} \right)\tilde \psi \left( \textbf{r} \right) = \frac{{\lambda _0^* }}{4}\psi \left( \textbf{r} \right),
\end{equation}
where the translated empirical pseudo-potential is defined by
\begin{equation}
\tilde U\left( \textbf{r} \right) =  - \frac{1}{4} \mathop \lambda \limits^ \to  \textbf{W}(\textbf{a}-<\textbf{a}>)= - \frac{1}{4}
\mathop \lambda \limits^ \to  {}^ * \textbf{W}\textbf{r} ,
\end{equation}
and the translated normalization Lagrange multiplier is now
\begin{equation}
\lambda_0^ *=\lambda_0+\mathop \lambda \limits^ \to  \textbf{W}\left\langle a \right\rangle
\end{equation}
It is noteworthy to mention that (28) is identical to the so-called \textit{zero-mean} form of the SWE employed in many works on FIM-based inference [56].
Eq. (23) is now re-cast as
\begin{equation}
\int {\tilde f\left( \textbf{r} \right)} \left[ {\left( {\frac{{d\ln \tilde f\left( \textbf{r} \right)}}{{d\textbf{r}}}} \right)^2  - 4\textbf{r}\frac{{d\tilde U\left( \textbf{r} \right)}}{{d\textbf{r}}}} \right]d\textbf{r} = 0.
\end{equation}
Note that
\begin{equation}
\int {\frac{1}{{f\left( \textbf{a} \right)}}\left( {\frac{{d f\left( \textbf{a} \right)}}{{d \textbf{a}}}} \right)^2 d\textbf{a}}= \left\langle {\textbf{a}\frac{{dU\left( \textbf{a} \right)}}{{d\textbf{a}}}} \right\rangle _{f\left( \textbf{a} \right)}  = \left\langle {\textbf{r}\frac{{dU\left( \textbf{r} \right)}}{{d\textbf{r}}}} \right\rangle _{\tilde f\left( \textbf{r} \right)}  =
\int {\frac{1}{{\tilde f\left( \textbf{r} \right)}}\left( {\frac{{d \tilde f\left( \textbf{r} \right)}}{{d \textbf{r}}}} \right)^2 d\textbf{r}},
\end{equation}
where, $\left\langle  \bullet  \right\rangle _{f\left(  \bullet  \right)}$ denotes the expectation evaluated with respect to $f(\bullet)$.
Substituting (30) into (32) and solving yields
\begin{equation}
\tilde f\left( \textbf{r} \right) = \exp \left[ { \mp \int {\sqrt { - \mathop \lambda \limits^ \to {}^ *    \textbf{W}\textbf{r}} } d\textbf{r}} \right].
\end{equation}
Setting
\begin{equation}
\mathop \lambda \limits^ \to {}^ *   =   - \frac{{\textbf{W}\textbf{r}}}{{\sigma ^4 }}
\end{equation}
results in the pdf
\begin{equation}
\tilde f\left( \textbf{r} \right)=  \exp \left[ { - \frac{{\textbf{W}\textbf{r}^T\textbf{r} }}{2\sigma^2}} \right]\\
\end{equation}
Thus
\begin{equation}
\begin{array}{l}
f\left( \textbf{a} \right) = \frac{{\exp \left[ { - \frac{{\textbf{W}\left( {\textbf{a} - \left\langle \textbf{a} \right\rangle } \right)^T \left( {\textbf{a} - \left\langle \textbf{a} \right\rangle } \right)}}{{2\sigma ^2 }}} \right]}}{{\tilde Z}} = \frac{{\exp \left[ { - \frac{{\sum\limits_{k = 1}^M {W_{ki} \sum\limits_{i = 1}^{N_c } {\left( {a_i  - \left\langle {a_i } \right\rangle } \right)^T \left( {a_i  - \left\langle {a_i } \right\rangle } \right)} } }}{{2\sigma ^2 }}} \right]}}{{\tilde Z}};\\
\\
\tilde Z = \int {\exp \left[ { - \frac{{\textbf{W}\left( {\textbf{a} - \left\langle \textbf{a} \right\rangle } \right)^T \left( {\textbf{a} - \left\langle \textbf{a} \right\rangle } \right)}}{{2\sigma ^2 }}} \right]d\textbf{a}} .
\end{array}
\end{equation}
Solving (37) yields
\begin{equation}
f\left( \textbf{a} \right) = \frac{1}{{\left( {2\pi \sigma ^2 } \right)^{\frac{{N_c }}{2}} }}\exp \left[ {\frac{{\left( {\textbf{v} - \textbf{W}\textbf{a}} \right)^T \left( {\textbf{v} - \textbf{W}\textbf{a}} \right)}}{{2\sigma ^2 }}} \right].
\end{equation}
From (28)
\begin{equation}
\left\langle {\textbf{r}^T \textbf{r}} \right\rangle  = \left\langle {\left( {\textbf{a} - \left\langle \textbf{a} \right\rangle } \right)^2 } \right\rangle  = \left\langle {\textbf{a}^T \textbf{a}} \right\rangle  - \left\langle \textbf{a} \right\rangle ^T \left\langle \textbf{a} \right\rangle  = \sigma ^2;
\end{equation}
With the aid of (28), (32), and (36), (37) yields the matrix FIM in the form [45]
\begin{equation}
I\left[ f \right] = \frac{{\textbf{W}^T \textbf{W}}}{{\sigma ^2 }}
\end{equation}
For normal distributions, the Cramer-Rao bound is always saturated [24, 45].  Thus, the diagonal covariance matrix is of the form
\begin{equation}
C = \sigma ^2(\textbf{W}^T\textbf{W})^{-1}.
\end{equation}

To formally establish the pseudo-inverse relation, (10), (17), and (35) yield
\begin{equation}
\begin{array}{l}
- \int {\textbf{W}^T \mathop \lambda \limits^ \to {}^ *  \psi ^2 \left( \textbf{a} \right)} d\textbf{a} = \textbf{W}^T \textbf{W}\left\langle a \right\rangle  = \textbf{W}^T \textbf{v} \\
 \end{array}
 \end{equation}
Thus
\begin{equation}
\left\langle \textbf{a} \right\rangle  = \left( {\textbf{W}^T \textbf{W}} \right)^{-1}\textbf{W}^T \textbf{v} = \textbf{W}^{\dagger}\textbf{v}.
\end{equation}
The pseudo-inverse condition (43) may be readily shown to be an \textit{efficient estimator} of \textbf{a}.
\section{Reciprocity relations and the Legendre transform structure}
The basic mathematical apparatus and theoretical considerations for deriving the reciprocity relations and the LTS have been established [28, 43].  Thus, only the pertinent FIM-relations  in a vector setting for least square constraints are to be  stated.
Multiplying (21) by $4\psi (\textbf{a})$ and integrating yields in vector form after re-arranging the terms
\begin{equation}
\textbf{I}\left[ \psi \right]=\lambda_0+ \mathop \lambda \limits^ \to \textbf{W}<\textbf{a}>
\end{equation}
To treat the component-wise case, the following definition is invoked
\begin{equation}
\int a_if(\textbf{a})d\textbf{a}=<a_i>,
\end{equation}
yielding
\begin{equation}
\textbf{I}\left[ \psi \right] = \lambda _0  + \sum\limits_{k = 1}^M {\lambda _k \sum\limits_{i = 1}^{N_c } {W_{ki} \left\langle {a_i } \right\rangle } }.
\end{equation}
Taking derivatives of (46) with respect to $\lambda_k$ results in
\begin{equation}
\frac{{\partial \textbf{I}\left[ \psi \right]}}{{\partial \lambda _k }} = \frac{{\partial \lambda _0 }}{{\partial \lambda _k }} + \sum\limits_{i = 1}^{N_c } {W_{ki} \left\langle {a_i } \right\rangle }  + \sum\limits_{\scriptstyle j = 1 \hfill \atop
  \scriptstyle j \ne k \hfill}^M {\lambda _j \frac{{\partial \sum\limits_{i = 1}^{N_c } {W_{ji} \left\langle {a_i } \right\rangle } }}{{\partial \lambda _k }}}.
\end{equation}
Specifying
\begin{equation}
\frac{{\partial \lambda _0 }}{{\partial \lambda _k }} =  - \sum\limits_{i = 1}^{N_c } {W_{ki} \left\langle {a_i } \right\rangle }
\end{equation}
in (47), yields the generalized Fisher-Euler theorem in a vector setting for least squares constraints
\begin{equation}
\frac{{\partial \textbf{I}\left[ \psi \right]}}{{\partial \lambda _k }} = \sum\limits_{\scriptstyle j = 1 \hfill \atop
  \scriptstyle j \ne k \hfill}^M {\lambda _j \frac{{\partial \sum\limits_{i = 1}^{N_c } {W_{ji} \left\langle {a_i } \right\rangle } }}{{\partial \lambda _k }}}.
  \end{equation}
Setting
\begin{equation}
\Theta _k \left( a_i \right) = \sum\limits_{i = 1}^{N_c } {W_{ki} a_i },
\end{equation}
With the aid of (45), (50) takes the form
\begin{equation}
\left\langle {\Theta _k \left( a_i \right)} \right\rangle  = \sum\limits_{i = 1}^{N_c } {W_{ki} \left\langle {a_i } \right\rangle }.
\end{equation}

With the aid of (21), (44),(45) and (50), the following relation is obtained
\begin{equation}
\begin{array}{l}
\lambda _0  + \sum\limits_{k = 1}^M {\lambda _k \left\langle {\Theta _k \left( {a_i } \right)} \right\rangle }  =  - \sum\limits_{k = 1}^M {\lambda _k \left\langle {a_i \frac{{d\Theta _k \left( {a_i } \right)}}{{da_i }}} \right\rangle }  =  - \sum\limits_{k = 1}^M {\lambda _k \left\langle {\Theta _k \left( {a_i } \right)} \right\rangle }  \\
  \Rightarrow \lambda _0  =  - 2\sum\limits_{k = 1}^M {\lambda _k \left\langle {\Theta _k \left( {a_i } \right)} \right\rangle }    \\
  \Rightarrow I\left[ \psi  \right] =  - \sum\limits_{k = 1}^M {\lambda _k \left\langle {\Theta _k \left( {a_i } \right)} \right\rangle }  \\
 \end{array}
\end{equation}
Taking derivatives of the third term in (52) yields
\begin{equation}
\frac{{\partial I\left[ \psi  \right]}}{{\partial \lambda _j }} =  - \left\langle {\Theta _j \left( {a_i } \right)} \right\rangle  - \sum\limits_{k = 1}^M {\lambda _k \frac{{\partial \left\langle {\Theta _k \left( {a_i } \right)} \right\rangle }}{{\partial \lambda _j }}}
\end{equation}
Consider the relation that underlies the basis for the LTS [26, 43]
\begin{equation}
\lambda _0 \left( {\lambda _1 ,...,\lambda _M } \right) = I\left( {\left\langle {\Theta _1 \left( {a_i } \right)} \right\rangle ,...,\left\langle {\Theta _M \left( {a_i } \right)} \right\rangle } \right) - \sum\limits_{k = 1}^M {\lambda _k \left\langle {\Theta _k \left( {a_i } \right)} \right\rangle }.
\end{equation}
Taking derivatives of (54) with respect to $
\left\langle {\Theta _j \left( {a_i } \right)} \right\rangle$ and comparing the ensuing results with  (53) yield the reciprocity relation
\begin{equation}
\frac{{\partial I\left( {\left\langle {\Theta _1 \left( {a_i } \right)} \right\rangle ,...,\left\langle {\Theta _M \left( {a_i } \right)} \right\rangle } \right)}}{{\partial \left\langle {\Theta _j \left( {a_i } \right)} \right\rangle }} = \lambda _j.
\end{equation}
Likewise, taking derivatives of (53) with respect to $\lambda_k$ yields the reciprocity relation (48).
Substituting (55) into (52) yields a linear PDE to infer the FIM without having to solve the vector independent Schr\"{o}dinger-like equation
\begin{equation}
I\left[ \psi  \right] =  - \sum\limits_{k = 1}^M {\left\langle {\Theta _k \left( {a_i } \right)} \right\rangle } \frac{{\partial I\left( {\left\langle {\Theta _1 \left( {a_i } \right)} \right\rangle ,...,\left\langle {\Theta _M \left( {a_i } \right)} \right\rangle } \right)}}{{\partial \left\langle {\Theta _j \left( {a_i } \right)} \right\rangle }}.
\end{equation}
Specifying
\begin{equation}
I\left[ \psi  \right] = \sum\limits_{k = 1}^M {I_k \left[ \psi  \right]}  =
\sum\limits_{k = 1}^M {\exp \left[ {g\left( {\left\langle {\Theta _k \left( {a_i } \right)} \right\rangle } \right)} \right]},
\end{equation}
and substituting (57) into (56) yields
\begin{equation}
I\left( {\left\langle {\Theta _1 \left( {a_i } \right)} \right\rangle ,...,\left\langle {\Theta _M \left( {a_i } \right)} \right\rangle } \right) = \sum\limits_{k = 1}^M {C_k |\left\langle {\Theta _k \left( {a_i } \right)} \right\rangle |^{-1} },
\end{equation}
where $C_k$ is a constant of integration.  Invoking (10) in (58) yields a candidate \textit{empirical FIM} for the modeling phase, defined solely in terms of the training data
\begin{equation}
I\left( {\left\langle {\Theta _1 \left( {a_i } \right)} \right\rangle ,...,\left\langle {\Theta _M \left( {a_i } \right)} \right\rangle } \right) = \sum\limits_{k = 1}^M {C_k v_k^{ - 1 } }.
\end{equation}
The utility and practical implementation is the task of ongoing work. Some of the potential implications of (59) are briefly discussed in Section 6.  Taking the derivative of (58) with respect to $\left\langle {\Theta _k \left( {a_i } \right)} \right\rangle$ yields
\begin{equation}
\frac{{\partial I\left( {\left\langle {\Theta _1 \left( {a_i } \right)} \right\rangle ,...,\left\langle {\Theta _M \left( {a_i } \right)} \right\rangle } \right)}}{{\partial \left\langle {\Theta _k \left( {a_i } \right)} \right\rangle }} =  - \frac{1 }{{\left\langle {\Theta _k \left( {a_i } \right)} \right\rangle }}C_k \left| {\left\langle {\Theta _k \left( {a_i } \right)} \right\rangle } \right|^{ - 1 }  =  - \frac{1 }{{\left\langle {\Theta _k \left( {a_i } \right)} \right\rangle }}I_k.
\end{equation}
Here, (60) describes a monotonically decreasing $I_k$ in the $\left\langle {\Theta _k \left( {a_i } \right)} \right\rangle$-direction.  Differentiating (60) yields
\begin{equation}
\frac{{\partial ^2 I\left( {\left\langle {\Theta _1 \left( {a_i } \right)} \right\rangle ,...,\left\langle {\Theta _M \left( {a_i } \right)} \right\rangle } \right)}}{{\partial \left\langle {\Theta _k \left( {a_i } \right)} \right\rangle \partial \left\langle {\Theta _l \left( {a_i } \right)} \right\rangle }} = 2 C_k \left| {\left\langle {\Theta _k \left( {a_i } \right)} \right\rangle } \right|^{ - 3} \delta _{kl},
\end{equation}
where $\delta _{kl}$ is the Kr\"{o}necker delta.  Here, (61) establishes the convexity of the FIM derived in this Section, thereby guaranteeing the existence of its inverse.
\section{Numerical examples}
This Section demonstrates the efficacy of the prediction model with the aid of the M-G DDE and two ECG signals.  The embedding dimension for all examples is evaluated using the false nearest neighbor method [57]    All numerical examples are evaluated for values of the forecasting time T= 1 and 5.  The largest positive Lyapunov exponent (LLE, hereafter)  is one of the simplest indicators of chaotic behavior.  From Refs. [17, 58, 59], it is evident that M-G DDE, with the delay time $\tau>14$ secs.,  possess a positive LLE.  Likewise, from [60,61], it has been established that the ECG signals comprising Record 207 of the MIT-BIH arrhythmia database possess a positive LLE, while [60] demonstrates that the signal comprising cudb/cu02 also has a positive LLE.
\subsection{Mackey-Glass equation}
The famed M-G DDE is described by
\begin{equation}
\frac{{dx\left( t \right)}}{{dt}} = \frac{{ax\left( t \right)\left( {t - \tau } \right)}}{{1 + x\left( t \right)^{10} \left( {t - \tau } \right)}} - bx\left( t \right),
\end{equation}
where, $a=0.2, b=0.1, \tau=30$ secs.  Here, $x(0)=1.2$.  Integration with a fourth-order Runge-Kutta routine yields the original solutions.  The value of the embedding dimension is $d=5$. The total number of points in the original solution is $1500$.  The procedures described in Sections 2 and 3 are then employed for $M=300 \in [0,300]$ secs. to obtain $<\textbf{a}>$, for $M_P=1494,~np=3,~N_c=56$.  These results and the concomitant MSE values are depicted in Figs. 1 and 2, respectively.

Here, Fig. 1 clearly demonstrates that the predicted results faithfully capture the dynamics embedded in the chaotic M-G time series'.  Fig. 2 expectedly demonstrates a slight distortion of the predicted signal vis-\'{a}-vis the original signal, as a consequence of long-term forecasting.  It may be argued that the number of coefficients $N_c=56$ is high and can forecast just about any signal.  This argument is not only tenuous at best for the case of chaotic signals, but is also orthogonal to the very reason causing the choosing of such a high value of $N_c$.  Specifically, Section 1 explicitly discusses the possible singularity (or near-singularity) of the embedding matrix \textbf{W}.  As stated therein,  large number of lags, \textbf{W} can result in volatile forecasts owing to ill-conditioning.  The M-G DDE in this example has a higher embedding dimension than other  prominent models (such as the Lorenz, H\'{e}non, etc.) (see, for eg. Ref. [3]), and thus the resulting \textbf{W} would be more prone to result in volatile forecasts.  As is evidenced by Figs. 1 and 2, this is not the case and the forecasts are clearly accurate and stable.

It is noteworthy to mention that the coefficients obtained from the training data during the modeling phase, which form the basis on which further prediction is done over a much larger time period and data sample size (as compared to those in the modeling phase), are unique to the specific data set under consideration.  Specifically, coefficients obtained from different data sets, for example $(i)$ the M-G DDE with a different value of the lag $\tau$ or $(ii)$ another nonlinear dynamics model, yield erroneous predictions if applied to a data set which differs from the one(s) they were obtained from.  This issue is the task on ongoing studies briefly described in Section 6 within the context of the results described in Sections 3 and 4, and will be presented elsewhere.
\subsection{MIT-BIH arrhythmia database Record 207}
This sub-Section employs a 300 secs. (5min.) ECG signal to demonstrate that the model described in this paper accurately predicts episodes (transients) which are the artifacts of a diseased heart over a reasonable period of time, even for a highly erractic/volatile signal.  The annotations are described in [62].  The signal is extracted from data obtained as a .mat data file from [63].  The sampling frequency of the data $\delta_s=360~Hz.$ [64],  the number of samples being $108,000$ for a total duration of $300~secs$.  The rationale for the choice of  300 secs. sample is to ensure that the portion of the signal, both during the modeling phase and the prediction phase that follows, possess sufficiently identifiable episodes which are documented in the reference annotations [65].  It would be desirable to conduct the study over the entire duration of the signal spanning around 30 mins.  However, this would yield simulation results that are visually incoherent, and hence the truncation of the signal length/duration.  The number of training samples from which the values of $<\textbf{a}>$ is obtained is $M=18,000$ $\in [0,50.0] $ secs. for forecasting times of  $T=1.0$ ($M_p=107,996)$ and $T=5.0$ ($M_p=107,995)$.  In all cases, $d=4$, $np=3$, and $N_c=35$.  All simulation results are presented for the MLII lead.    The simulation results depicting the predicted ECG signal superimposed over the original Record 207 signal for the MLII lead for T=1.0 and 5.0 respectively and the  concomitant MSE's are presented in Figs. 3 and 5, respectively.

Both Figs. (3) and (5) demonstrate that even for a relatively small duration of the modeling phase which comprises 1/6-th of the duration of the entire prediction exercise comprising of both the modeling phase and the prediction of new data, the prediction results are of a high quality.  The case corresponding to $T=5.0$ shows greater "undershoots" and "overshoots" of the peaks of the signal as compared to the case of $T=1.0$.  This degradation of prediction performance is expected.   \textit{The results of the MSE's for both cases of the forecasting time shows divergent peaks. These have been analyzed and found to be the result of the highly  erractic/volatile nature of the signal.  It is important to note that they do not constitute any volatility in the prediction since the profiles of the predicted signals are demonstrated to be very much in accord with the original signal.  Further, in both cases, following every divergent peak which can even be visually related to erratic signal quality in Figs. (3) and (5), the prediction returns to "normalcy" which is defined by a low MSE value.  This would not be the case of a volatile prediction cased by ill-conditioning or any other factor (see Section 1), because the errors cased by volatile predictions tend to cascade}.

Figs. (4) and (6) focus on specific regions of interest.  Figs. 4(a) and 6(a) depict the overlaid overlaid and predicted signals of the modeling phase for the cases of the forecasting time $T=1.0$ and $T=5.0$, respectively.  To establish the accuracy of the FIM based model, the prediction of the episodes corresponding to the various conditions of the diseased heart documented in the reference annotations [65], that can be visually determined in the period $[0,300]$ secs. are demonstrated in Figs. 4(b)-(d) and 6(b)-(d) for $T=1.0$ and $T=5.0$, respectively.  On inspection of Figs. 4(b) and 6(b), $(i)$ the instance of ventricular tachycardia identified by "+" and defined by "(VT" in the reference annotations at 38.522 secs., immediately followed by three instances of premature ventricular contraction identified by "V" and $(ii)$ the onset of ventricular flutter/fibrillation identified by "[" at 40.736 secs. in the reference annotation followed by an instance of ventricular flutter identified by "+" and defined by "(VFL" in the reference annotations at 40.803 secs. and the subsequent termination of the ventricular flutter/fibrillation identified by "]" at 50.972 secs. in the reference annotation can be easily identified.  This region is of particular importance since it spans \textit{both} the modeling phase from which the \textit{working hypothesis} is determined from the training data, \textit{and} the prediction of new data values.

On inspection of Figs. 4(c) and 6(c), $(i)$ the onset of ventricular flutter/fibrillation identified by "[" at 54.764 secs. in the reference annotation followed by an instance of ventricular flutter identified by "+" and defined by "(VFL" in the reference annotations at 54.869 secs. and the subsequent termination of the ventricular flutter/fibrillation identified by "]" at 50.972 secs. in the reference annotation and $(ii)$ the instance of ventricular tachycardia identified by "+" and defined by "(VT" in the reference annotations at 61.839 secs. (1:01.839 mins.), immediately followed by three instances of premature ventricular contraction identified by "V", can be easily identified.  Finally, on inspection of Figs. 4(d) and 6(d), the onset of ventricular flutter/fibrillation identified by "[" at  269.467  secs. (4:29.467 mins.) in the reference annotation followed by an instance of ventricular flutter identified by "+" and defined by "(VFL" in the reference annotations at 129.586 secs. (4:29.586 mins.) and the subsequent termination of the ventricular flutter/fibrillation identified by "]" at 240.906  secs. (4:40.906 mins.) in the reference annotation, can be easily identified.  In all cases depicted in Figs. (4) and (6), it is observed that the quality of the prediction is high, with the case of the example with $T=5.0$ being marginally degraded vis-\'{a}-vis the case with $T=1.0$, which is expected.
\subsection{Creighton University VTA database Record cudb/cu02}
This sub-Section demonstrates the robustness of model described in this paper for an extended ECG signal, even for a highly erractic/volatile signal displaying the symptoms of cardiac VTA.   The signal is extracted from data obtained as a .mat data file from [66].  The sampling frequency of the data $\delta_s=250~Hz.$, and the number of samples is $127,232$, for a total duration of $8:28.928~mins.$ [67].  In order to study the robustness of the prediction performance of the FIM-based model, the number of training samples from which the values of $<\textbf{a}>$ is obtained is chosen to be $M=30,000$ $\in [0,120.0] $ secs. for  $T=1.0$ ($M_p=127,228)$ and $T=5.0$ ($M_p=127,227)$.  In all cases, $d=4$, $np=3$, and $N_c=35$.

The high quality of the model is attested by the fidelity of the predicted ECG profiles with the original signals, and, the MSE's.   The simulation results depicting the predicted ECG signal superimposed over the original signal for T=1.0 and 5.0 and the  concomitant MSE's, are presented in Figs. 7 and 8, respectively.   Similar to the case described in Section 5.2, the results of the MSE's for both forecasting times depict divergent peaks. These have been analyzed and found to be the result of the nature of the signal.  Again, note that these divergent peaks in the MSE's do not constitute any volatility in the prediction since the profiles of the predicted signals are demonstrated to be very much in accord with the original signal.  Further, in both cases, following every divergent peak which can even be visually related to erratic signal quality in Figs. (7) and (8), the prediction returns to "normalcy" (defined by a low MSE value).  This would not be the case in a volatile prediction cased by ill-conditioning or any other factor (see Section 1), because the errors cased by volatility of the forecasting tend to cascade.

\section{Summary and conclusions}
A convenient framework for the modeling and forecasting of time series has been developed  within the ambit of a FIM-based inference model.  The modeling phase from which the \textit{working hypothesis} is derived has been provided with a QM connotation, by the formulation of a time independent Schr\"{o}dinger-like equation in a vector setting, employing least squares constraints.  This has been achieved by invoking the MFI principle. Apart from the obvious theoretical implications, this allows for the systematic derivation and categorization of the \textit{working hypothesis} and the subsequential forecasting phase.  The pdf and the pseudo-inverse relations have been self-consistently inferred by invoking the QM virial theorem for the case of normal distributions.    The reciprocity relations and the LTS for the modeling phase have been derived for the modeling phase.  This results in an intriguing form of the FIM for the modeling phase, which defines the \textit{working hypothesis}, described solely in terms of the observed data (the \textit{empirical FIM}).

The possible utilities of this form of the FIM are to derive principled expressions and values for the statistical dispersion.  This FIM expression  has no equivalent in prior MaxEnt models [7-10], which treat the statistical dispersion as an ad-hoc scaling [15].  The FIM-based model has been numerically tested, and its efficacy proven for the Mackey-Glass DDE, the ECG signal for the MLII lead from Record 207 of the MIT-BIH cardiac arrhythmia database, and the  ECG from Record cudb/cuo2 of the Creighton University VTA database.  The forecasting is consistently demonstrated to be of very high quality, and does not suffer from any signs of volatility caused by ill-conditioning of the embedding matrix \textbf{W} or any other factor.  The numerical experiments on the ECG demonstrate that the model presented in this paper is able to forecast salient episodes documented in the reference annotations for the said signals, to a very high degree of accuracy.  Ongoing work is focused on a two-pronged approach.  First, the \textit{empirical FIM} derived in Section 4 has been investigated within the context of its relationship to the FIM employed in Section 3, and the subsequent effects on the forecasting.  A fiduciary time dependence is induced into the modeling phase via a sliding window analysis.  This allows for the quality of prediction to be related to fundamental results governing the FIM, viz. the \textit{I-theorem} (the Fisher-equivalent of the H-theorem) [24], for both normal and non-equilibrium distributions.  Next, a principled comparison of the FIM-based model with existing nonparametric prediction models [12, 13] is in progress.  These works will be presented elsewhere.
\section*{Appendix A: Derivation of Eq. (16)}
 \renewcommand{\theequation}{A.\arabic{equation}}
  \setcounter{equation}{0}  
  For the Fisher information matrix $\left[ F \right]$, each element is defined by
 \begin{equation}
F_{ij}  = \int {f\left( \textbf{a} \right)} \left[ {\frac{{\partial \ln f\left( \textbf{a} \right)}}{{\partial a_i }}\frac{{\partial \ln f\left( \textbf{a} \right)}}{{\partial a_j }}} \right]da.
\end{equation}
For \textbf{a} possessing \textit{a-priori} iid entries
\begin{equation}
f\left( \textbf{a} \right) = \prod\limits_i {f_i \left( {a_i } \right)};\frac{{\partial \ln f\left( \textbf{a} \right)}}{{\partial a_i }} = \frac{{\partial \ln f_i \left( {a_i } \right)}}{{\partial a_i }}.
\end{equation}
Substituting (A.2) into (A.1) yields
\begin{equation}
F_{ij}  = \int {\prod\limits_k {f_k \left( {a_k } \right)} } \left[ {\frac{{\partial \ln f_i \left( {a_i } \right)}}{{\partial a_i }}\frac{{\partial \ln f_j \left( {a{}_j} \right)}}{{\partial a_j }}} \right]da_k
\end{equation}
For $i\ne j$
\begin{equation}
\begin{array}{l}
F_{ij}  = F_i F_j ;
F_i  = \int {f_i \left( {a_i } \right)\frac{{\partial \ln f_i \left( {a_i } \right)}}{{\partial a_i }}da_i }  = \frac{\partial }{{\partial a_i }}\int_I {f_i \left( {a_i } \right)da_i }  = \frac{\partial }{{\partial a_i }}\int 1  = 0,
\end{array}
\end{equation}
since all other integrals $da_k$ integrate to unity because of normalization.  For $i=j$,
\begin{equation}
F_{ii}  = \int {f_i \left( {a_i } \right)\left( {\frac{{\partial f_i \left( {a_i } \right)}}{{\partial a_i }}} \right)^2 da_i }.
\end{equation}
Thus, $\left[ F \right]$ is a diagonal matrix with each element defined by (A.5).  Note that $F_{ii}=I_i$, as used in Eq. (16).

\newpage
\begin{figure}[ht]
 \centering
 \subfigure[Predicted vs. original signals.]{
  \includegraphics[width =0.47\linewidth]{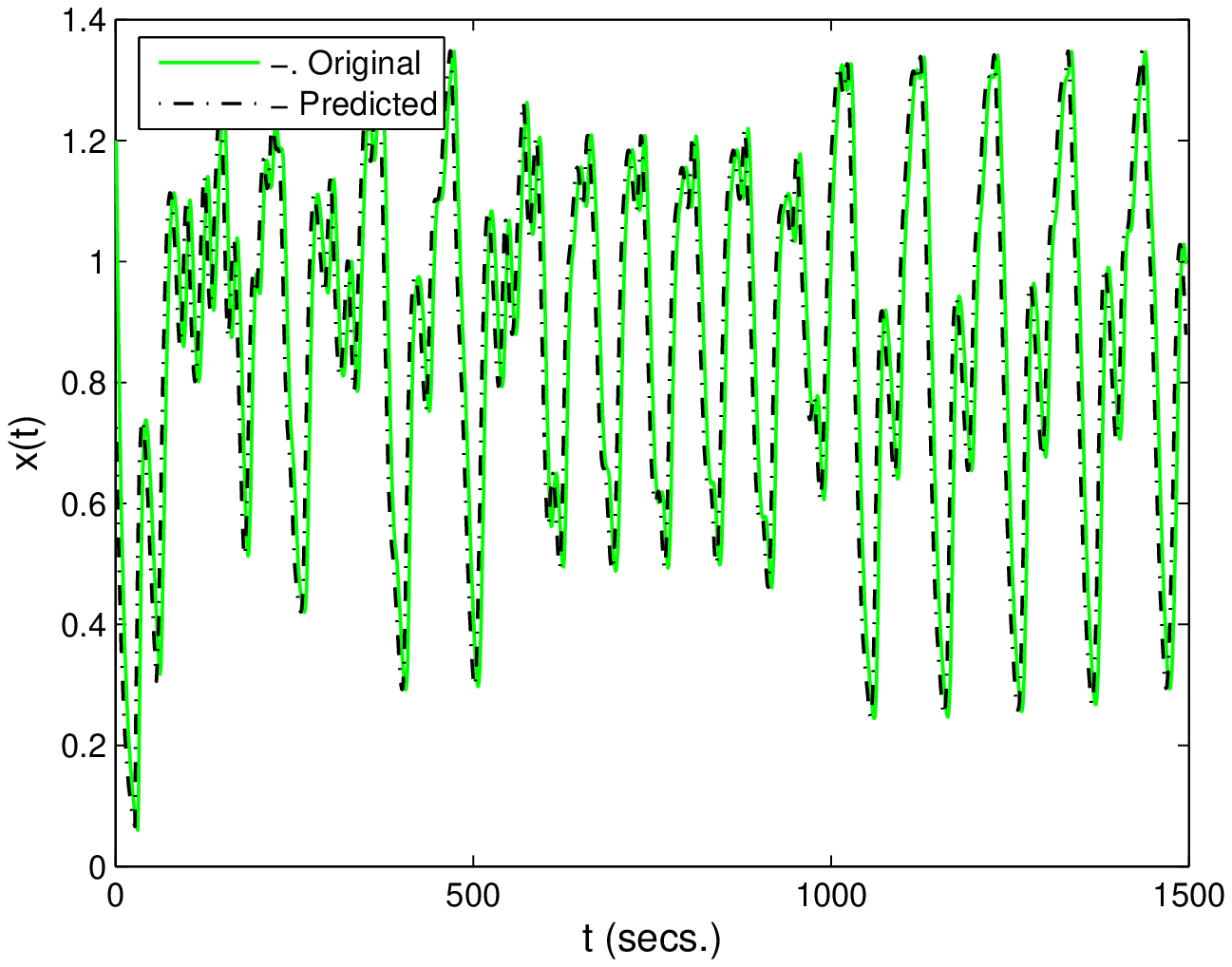}
   \label{fig:subfig1}
   }
\subfigure[MSE]{
  \includegraphics[width =0.47\linewidth]{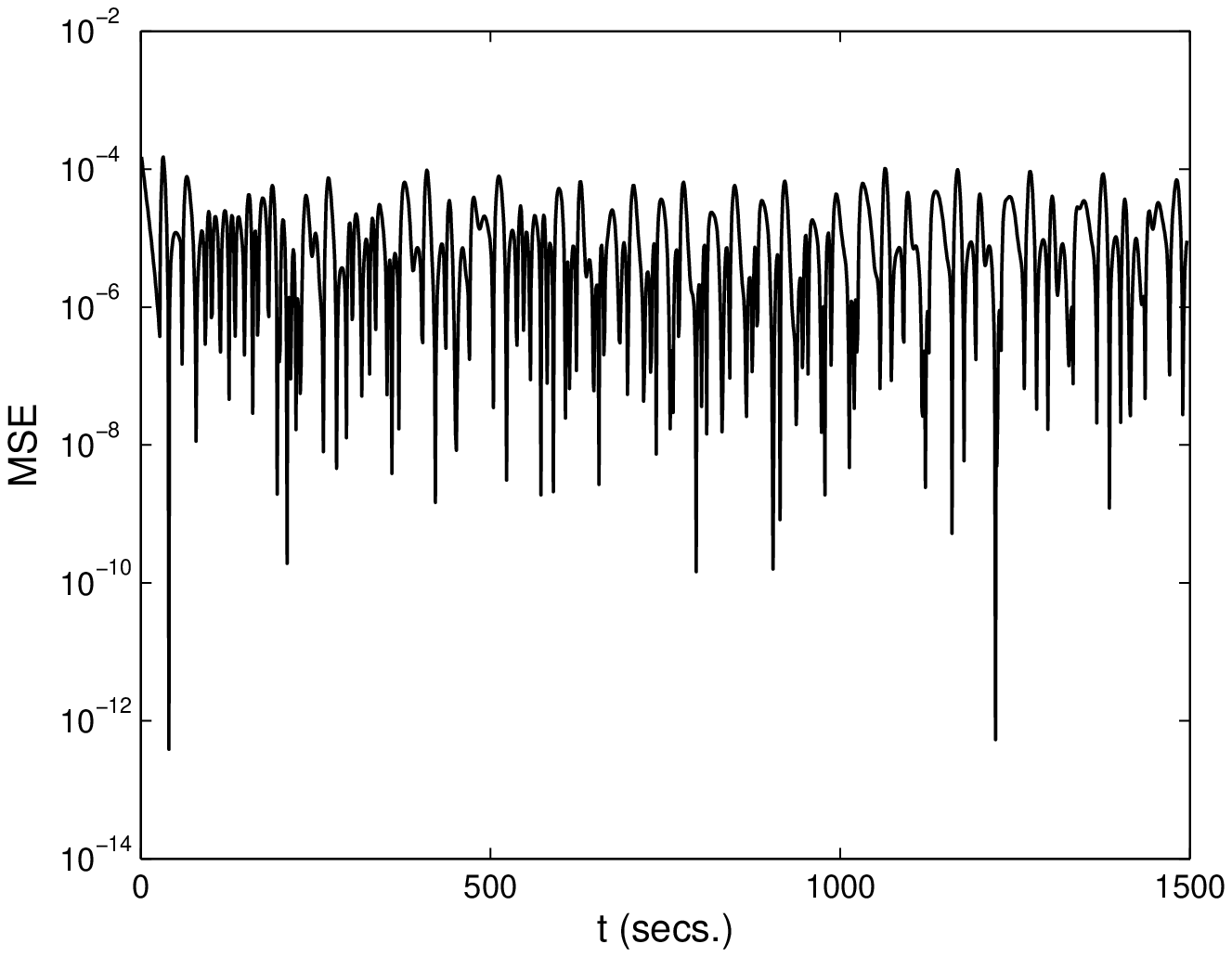}
   \label{fig:subfig1}
   }

\label{fig:subfigureExample}
 \caption[Optional caption for list of figures]{%
  Predicted vs. original signals and MSE for the Mackey-Glass equation, $\tau=30$ secs., d=5, T=1.0, and $M_p$=1494.  Modeling phase is $\in[0,300]$ secs. for M=300 training data. New data predicted $\in[300,1495]$ secs. is $M_p-M$=1195.  }
\end{figure}

\begin{figure}[ht]
 \centering
 \subfigure[Predicted vs. original signals.]{
  \includegraphics[width =0.47\linewidth]{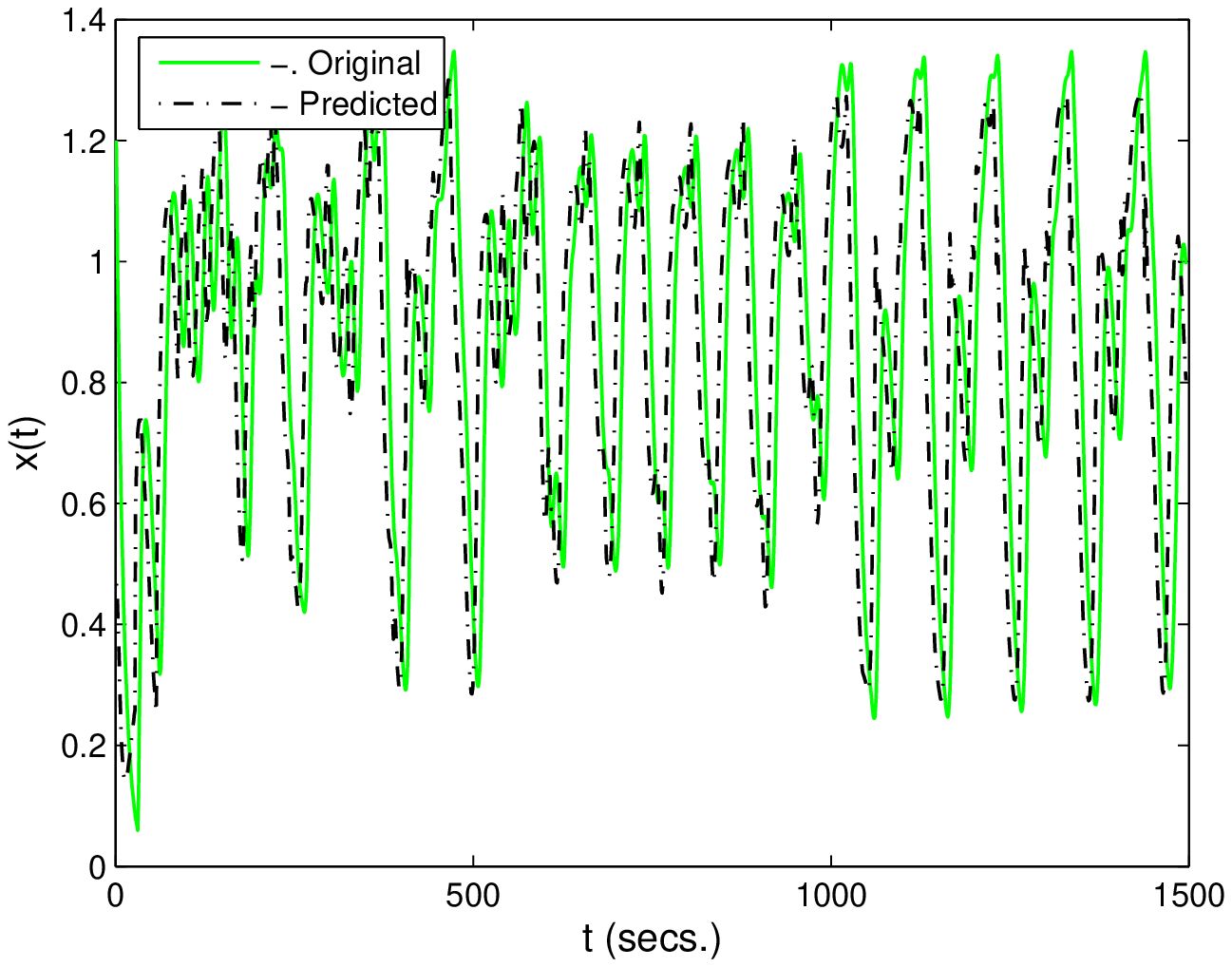}
   \label{fig:subfig1}
   }
 \subfigure[MSE]{
  \includegraphics[width =0.47\linewidth]{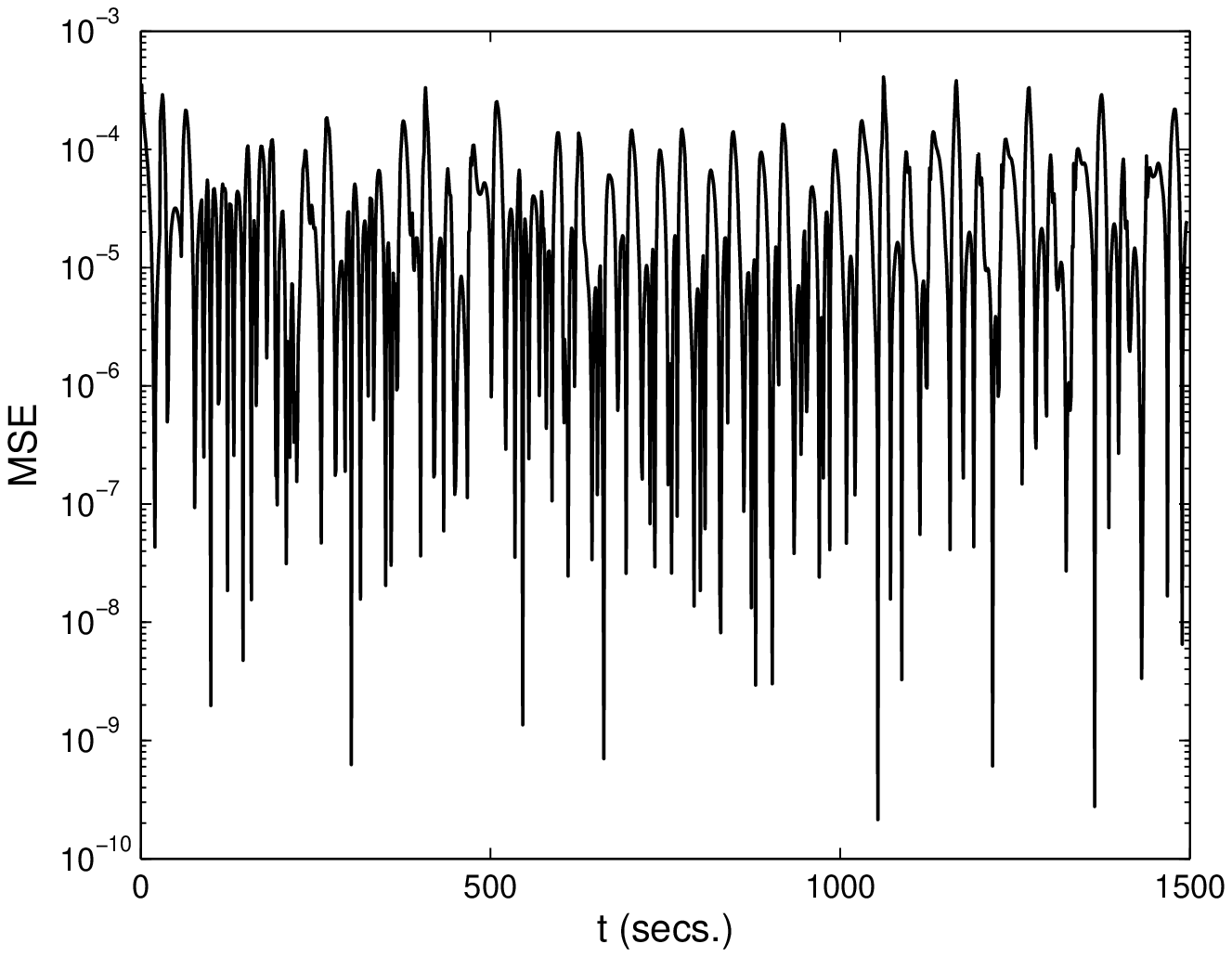}
   \label{fig:subfig1}
   }

\label{fig:subfigureExample}
 \caption[Optional caption for list of figures]{%
  Predicted vs. original signals and MSE for the Mackey-Glass equation, $\tau=30$ secs., d=5, T=5.0, and $M_p$=1494.  Modeling phase is $\in[0,300]$ secs. for M=300 training data. New data predicted $\in[300,1495]$ secs. is $M_p-M$=1195.   }
\end{figure}
\newpage
\begin{figure}[ht]
 \centering
 \subfigure[Predicted vs. original signals.]{
  \includegraphics[width =1.0\linewidth]{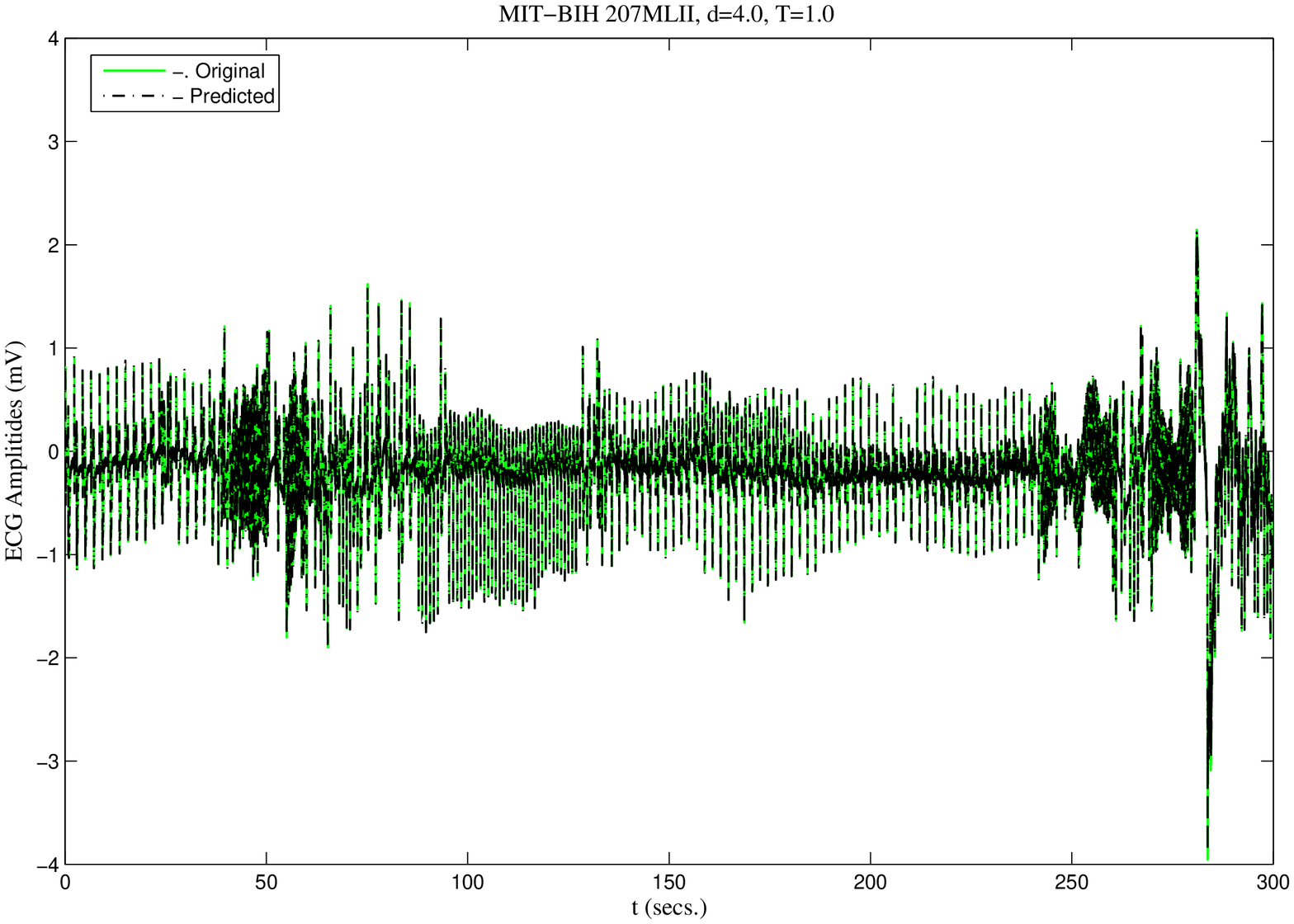}
   \label{fig:subfig1}
   }
 \subfigure[MSE]{
  \includegraphics[width =1.0\linewidth]{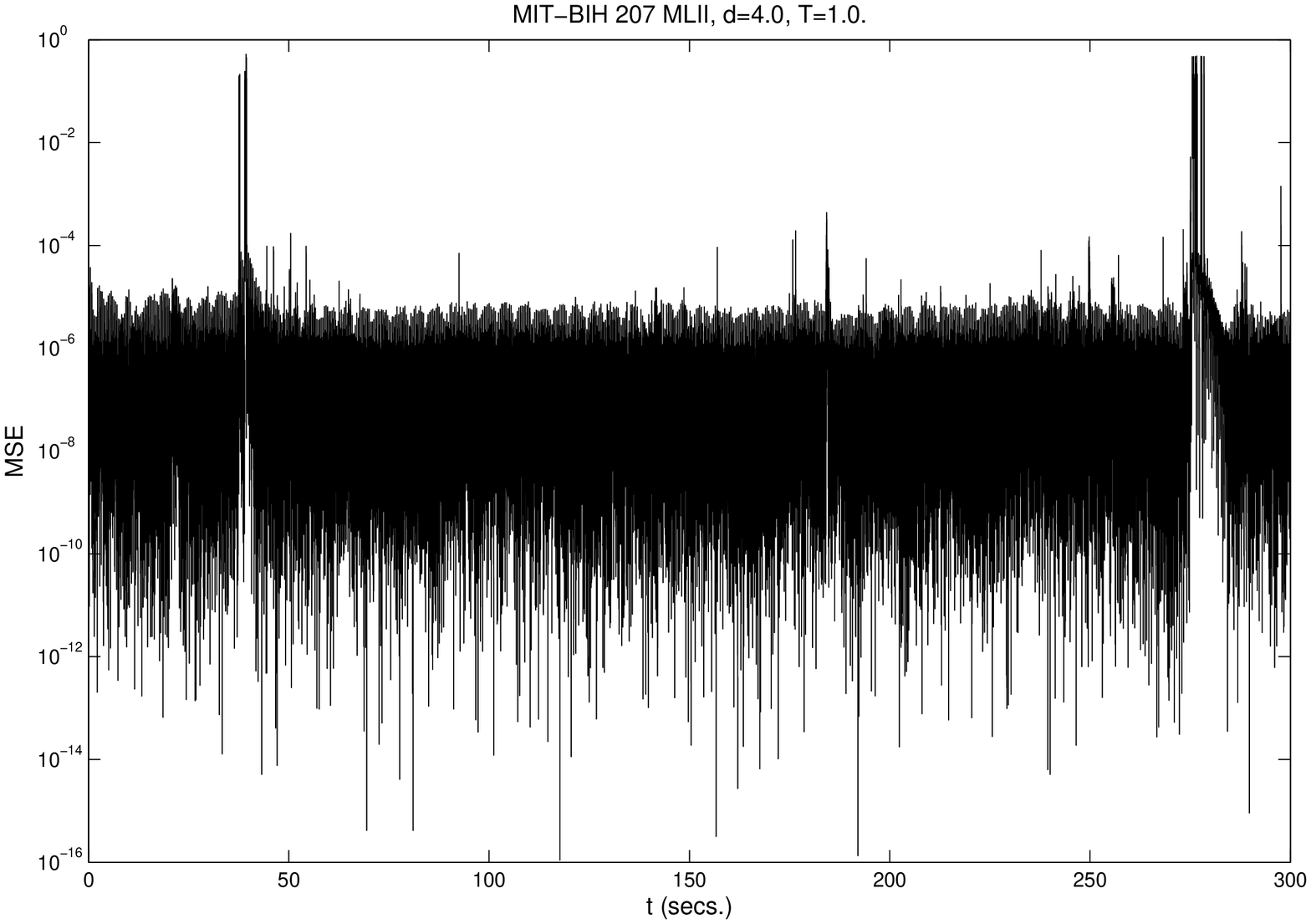}
   \label{fig:subfig1}
   }

\label{fig:subfigureExample}
 \caption[Optional caption for list of figures]{%
 Predicted vs. original signals and MSE for ECG signal from MIT-BIH arrhythmia database for Sample 207 and lead MLII, d=4, T=1.0, and $M_p=107, 996$.  Modeling phase is $\in[0,50]$ secs. for M=18,000 training data. New data predicted $\in[50,300]$ secs. is $M_p-M$=89,996.   }
\end{figure}
\newpage

\begin{figure}[ht]
 \centering
 \subfigure[Predicted vs. original signals.]{
  \includegraphics[width =0.47\linewidth]{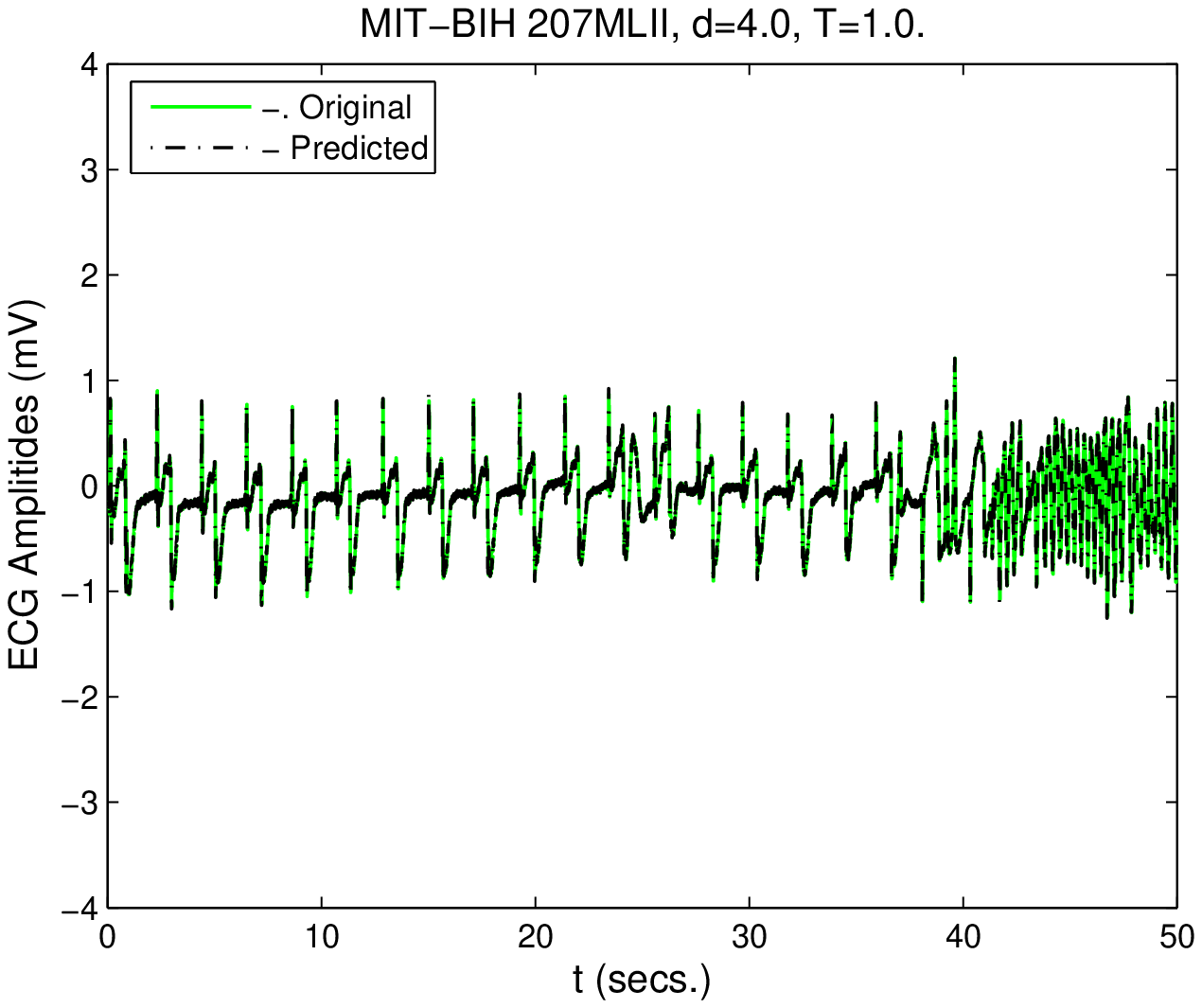}
   \label{fig:subfig1}
   }
 \subfigure[Predicted vs. original signals.]{
  \includegraphics[width =0.47\linewidth]{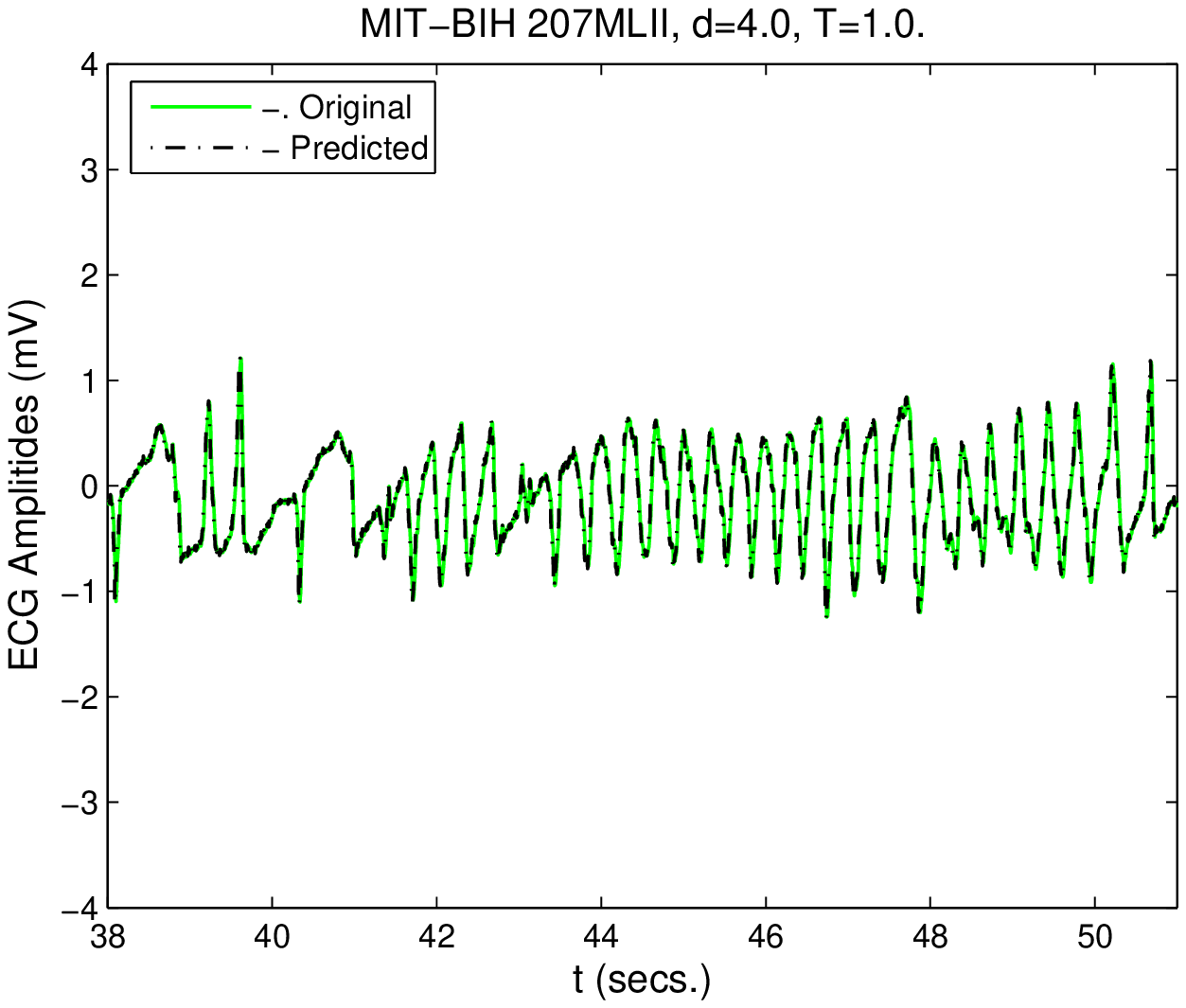}
   \label{fig:subfig1}
   }
 \subfigure[Predicted vs. original signals.]{
  \includegraphics[width =0.47\linewidth]{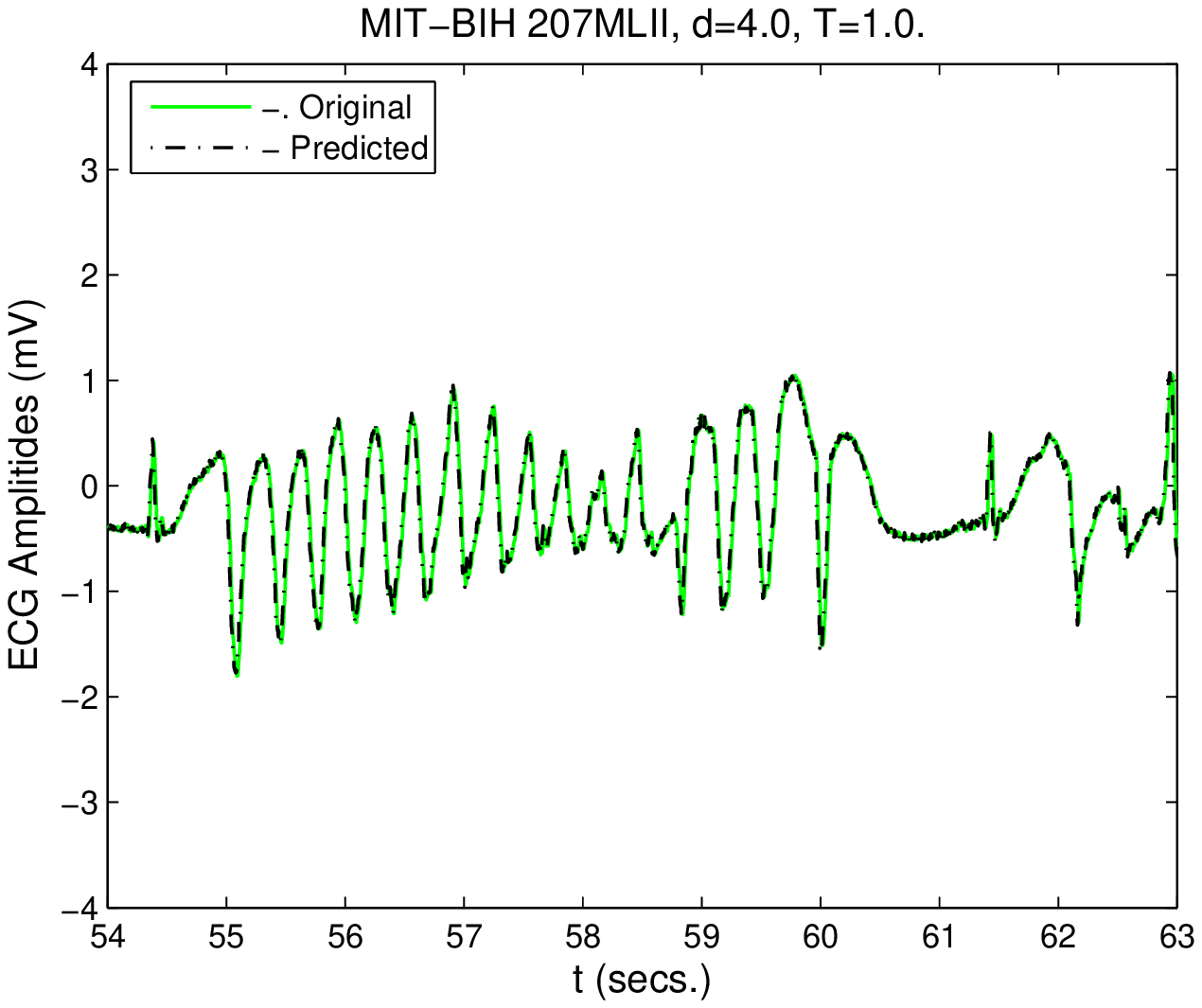}
   \label{fig:subfig1}
   }
\subfigure[Predicted vs. original signals.]{
  \includegraphics[width =0.47\linewidth]{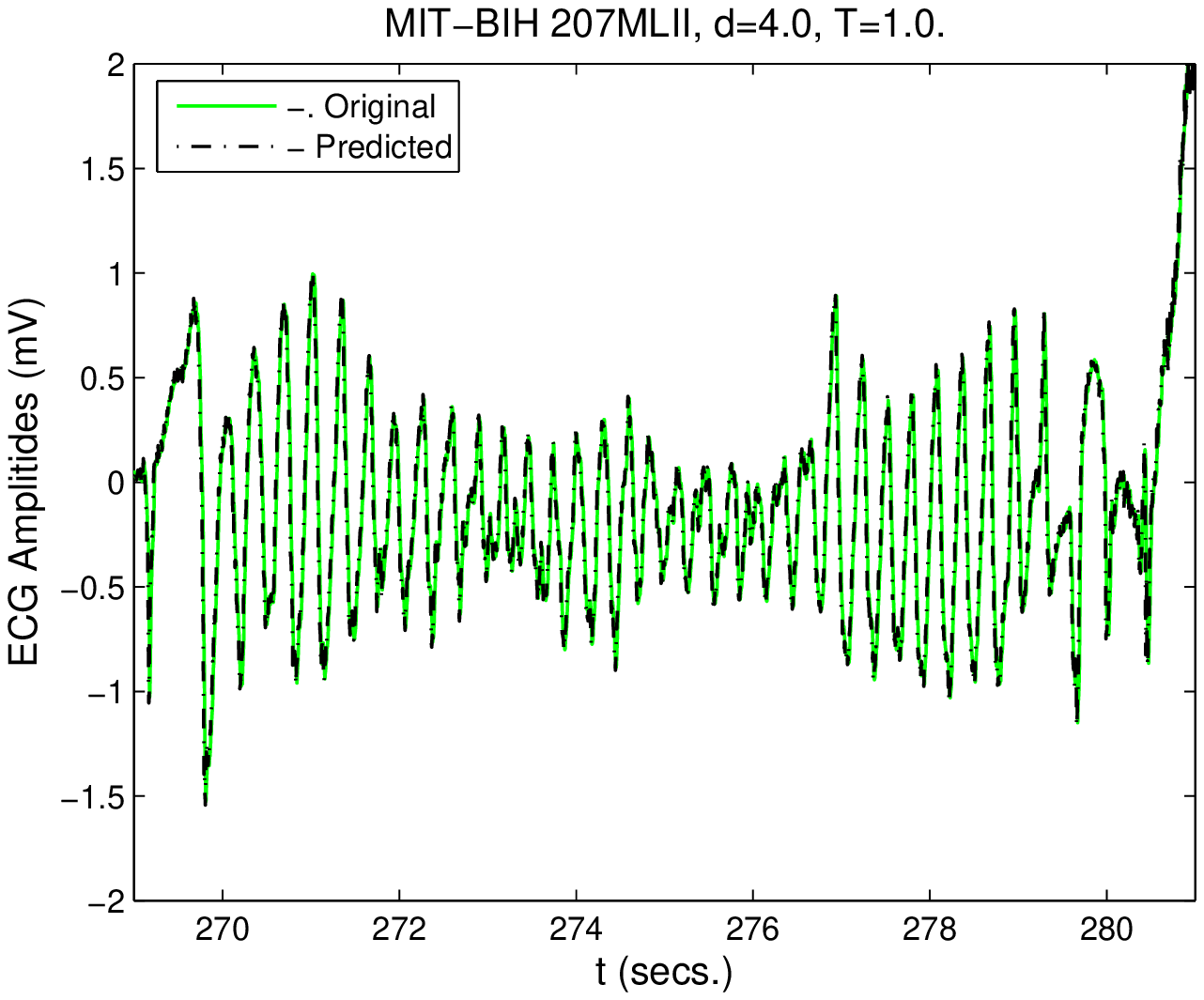}
   \label{fig:subfig1}
   }
 \caption[Optional caption for list of figures]{%
 Sample segments of predicted vs. original signals and MSE for ECG signal from MIT-BIH arrhythmia database for Sample 207 and lead MLII, d=4, T=1.0, and $M_p=107, 996$.  Modeling phase is $\in[0,50]$ secs. for M=18,000 training data. New data predicted $\in[50,300]$ secs. is $M_p-M$=89,996.    }
\end{figure}
\newpage
\begin{figure}[ht]
 \centering
 \subfigure[Predicted vs. original signals.]{
  \includegraphics[width =1.0\linewidth]{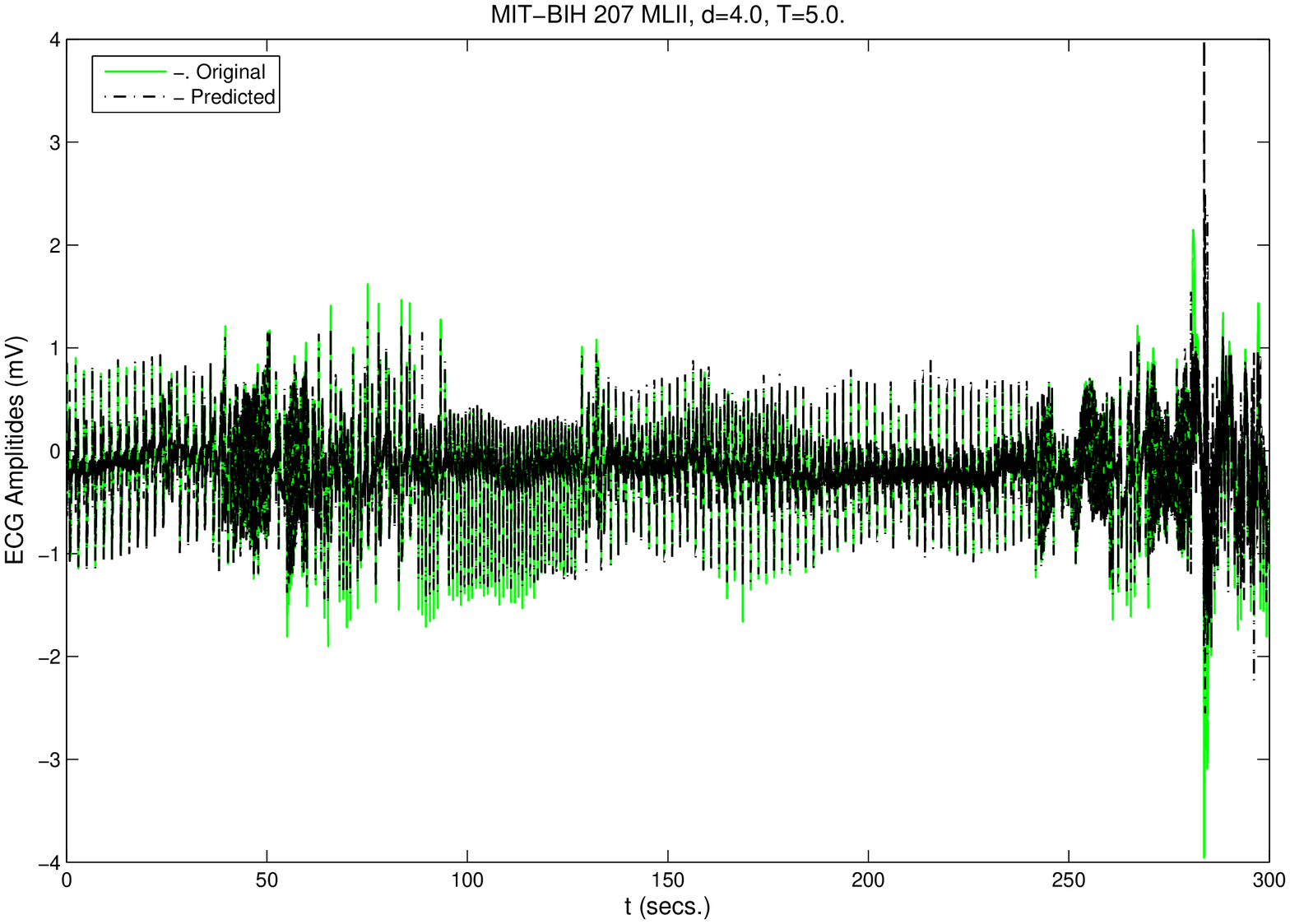}
   \label{fig:subfig1}
   }
 \subfigure[MSE]{
  \includegraphics[width =1.0\linewidth]{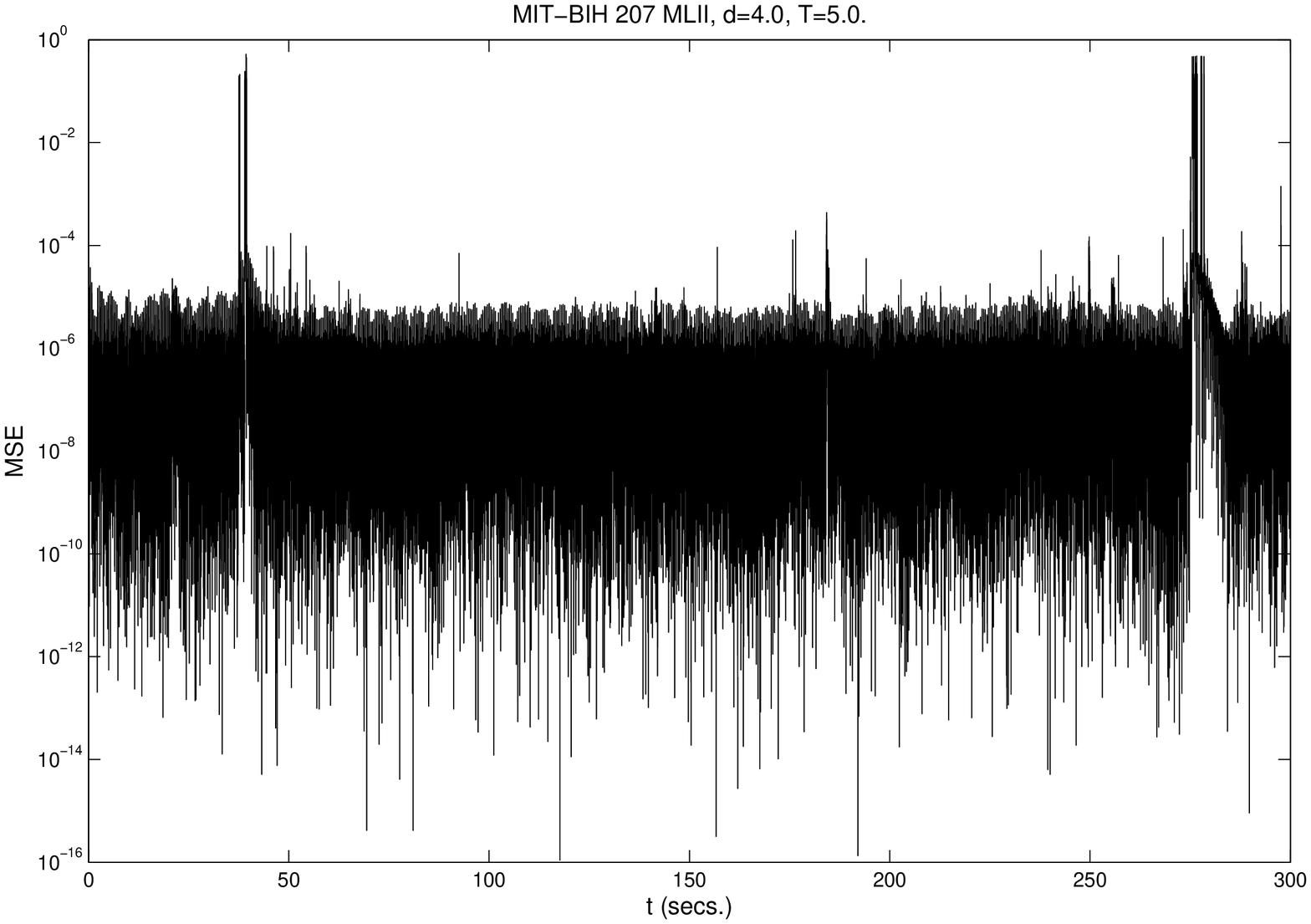}
   \label{fig:subfig1}
   }

\label{fig:subfigureExample}
 \caption[Optional caption for list of figures]{%
  Predicted vs. original signals and MSE for ECG signal from MIT-BIH arrhythmia database for Sample 207 and lead MLII, d=4, T=5.0, and $M_p=107, 995$.  Modeling phase is $\in[0,50]$ secs. for M=18,000 training data. New data predicted $\in[50,300]$ secs. is $M_p-M$=89,995.  }
\end{figure}
\newpage

\begin{figure}[ht]
 \centering
 \subfigure[Predicted vs. original signals.]{
  \includegraphics[width =0.47\linewidth]{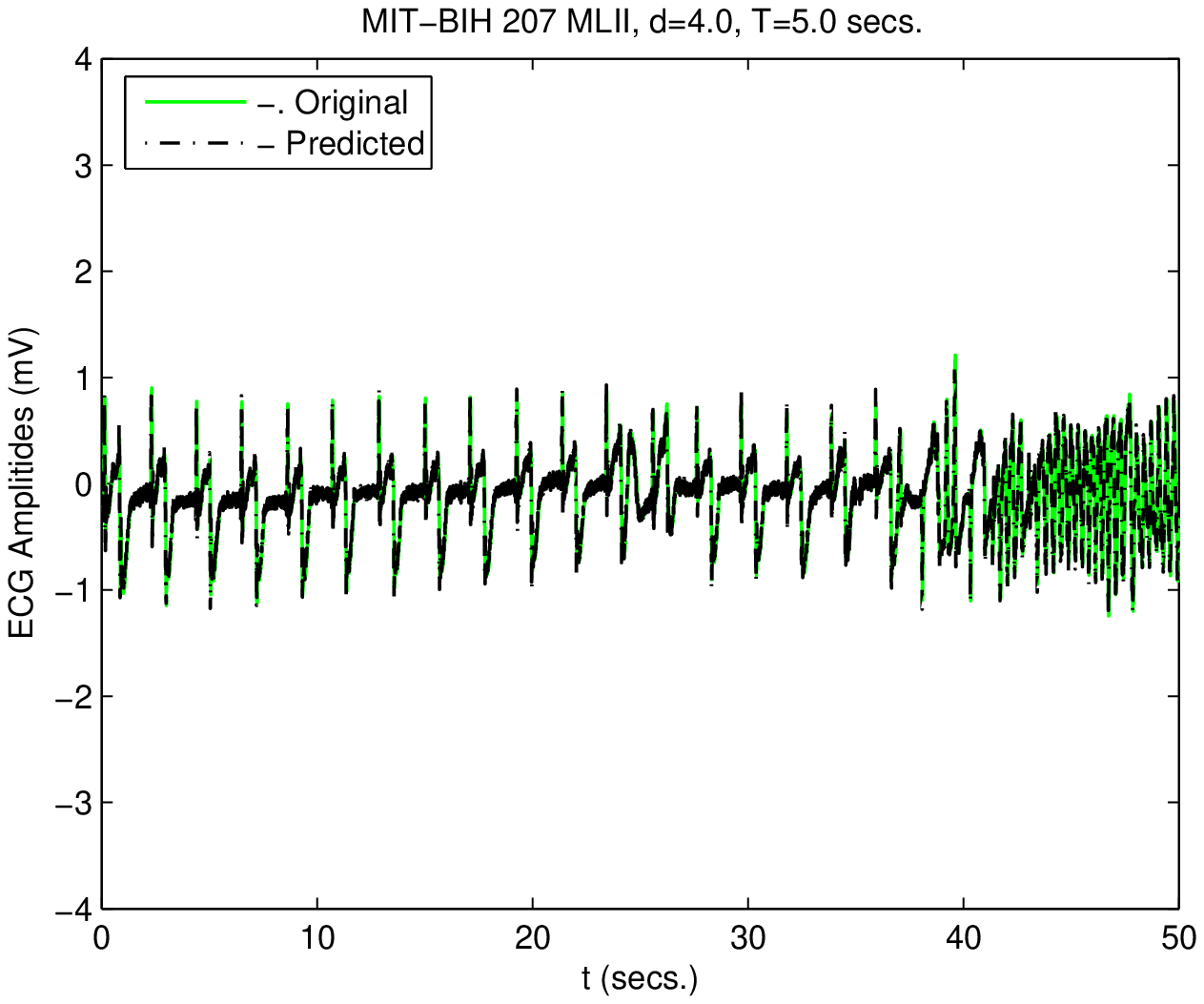}
   \label{fig:subfig1}
   }
 \subfigure[Predicted vs. original signals.]{
  \includegraphics[width =0.47\linewidth]{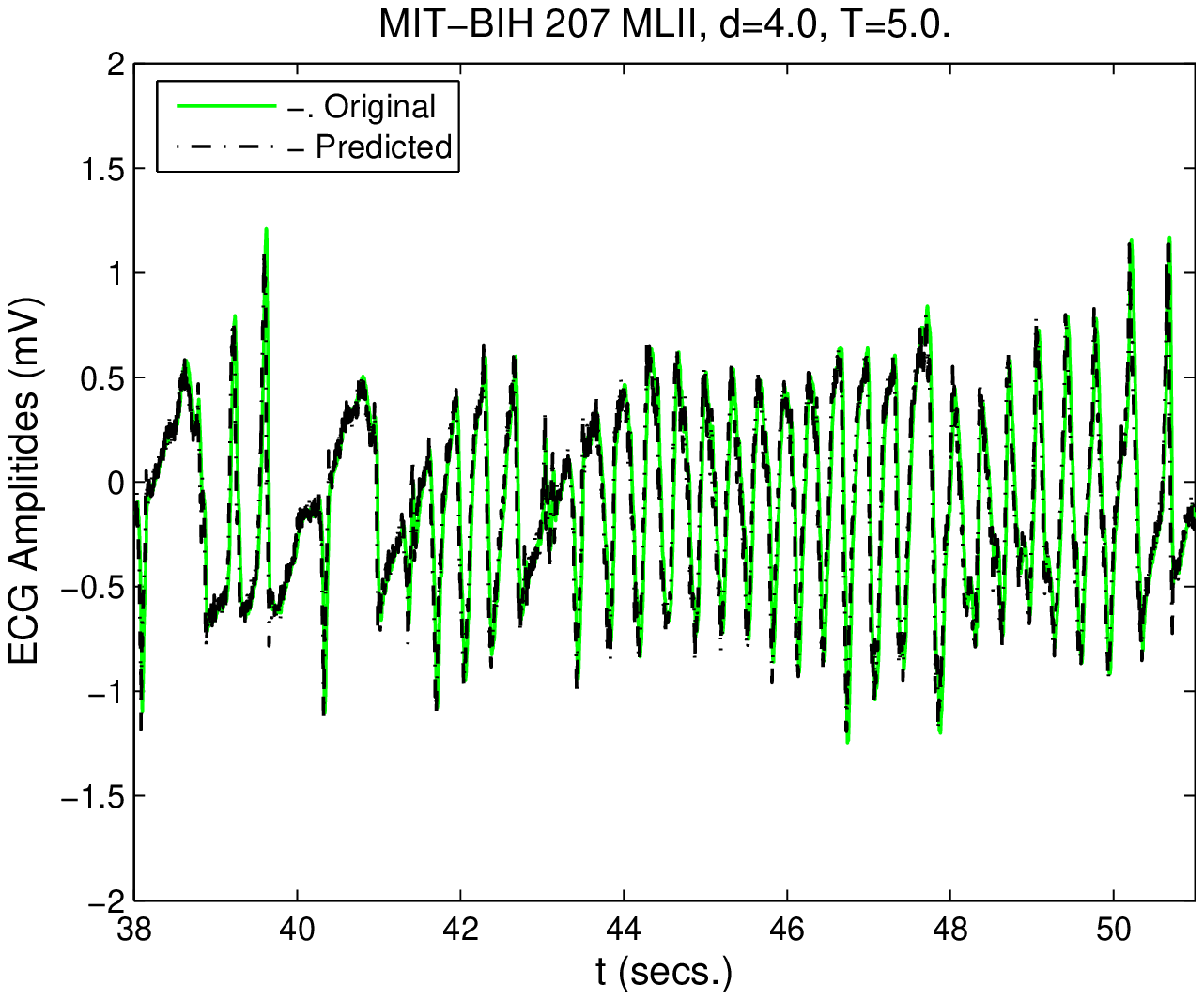}
   \label{fig:subfig1}
   }

 \subfigure[Predicted vs. original signals.]{
  \includegraphics[width =0.47\linewidth]{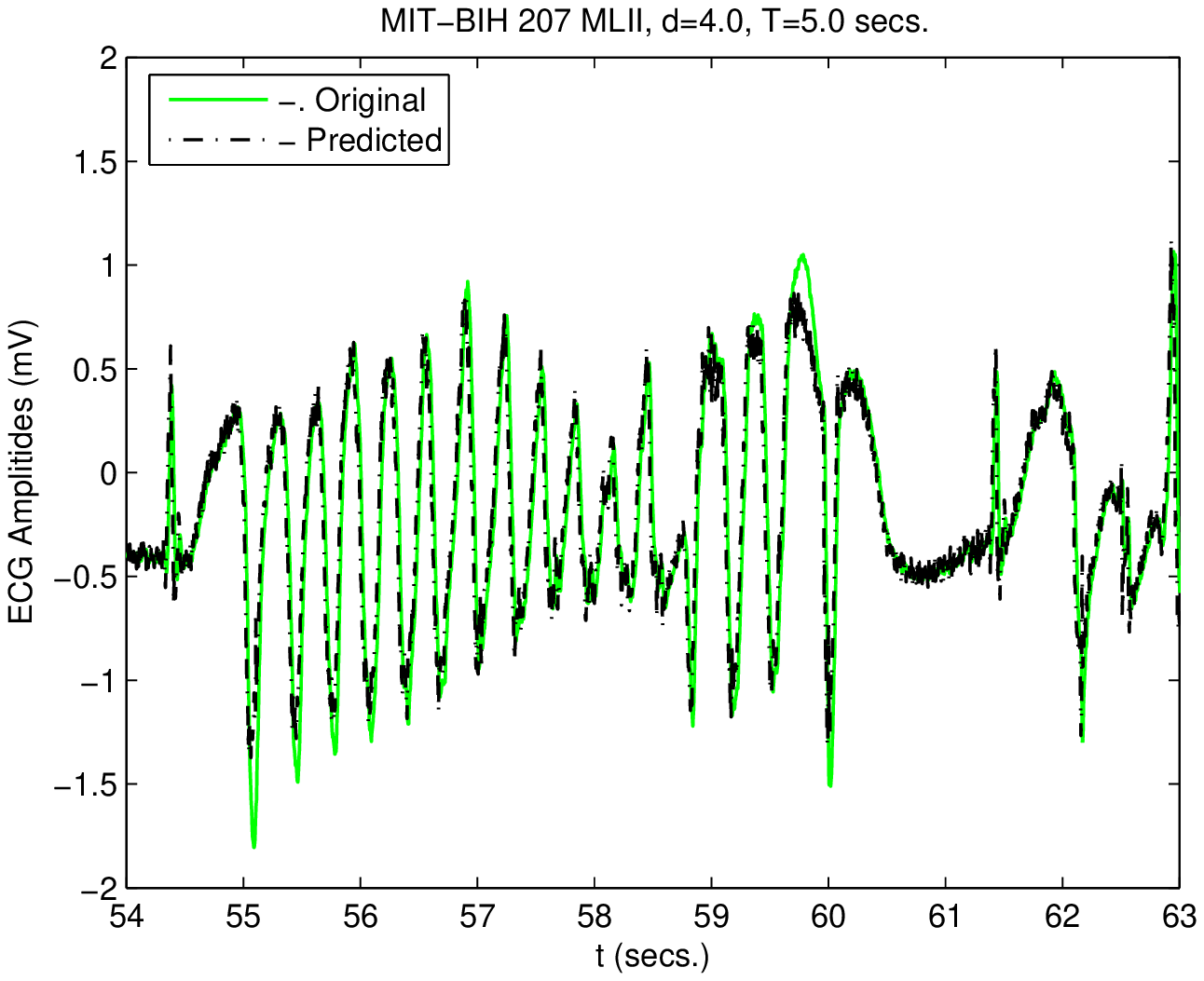}
   \label{fig:subfig1}
   }
   \subfigure[Predicted vs. original signals.]{
  \includegraphics[width =0.47\linewidth]{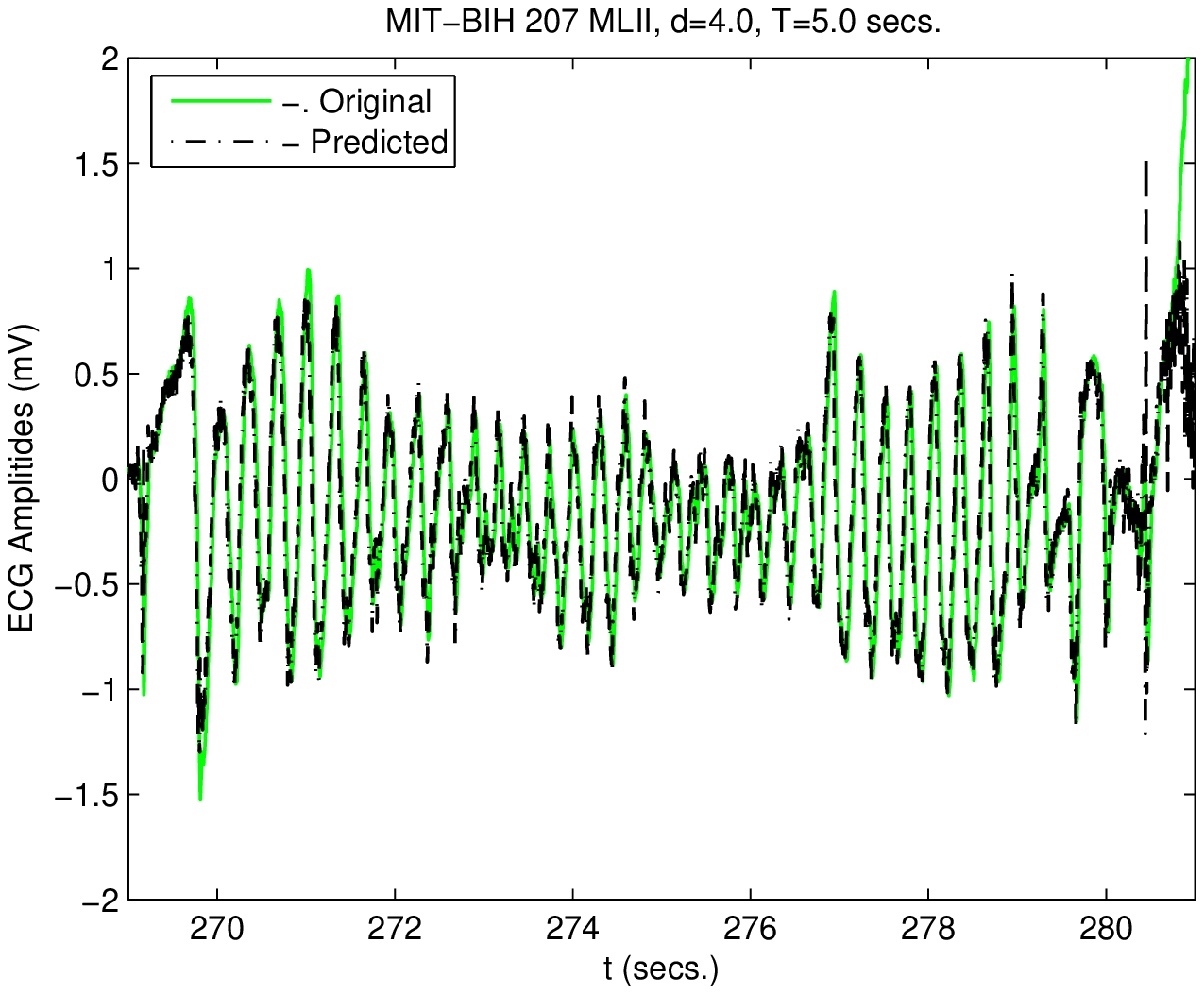}
   \label{fig:subfig1}
   }
\label{fig:subfigureExample}
 \caption[Optional caption for list of figures]{%
  Sample segments of predicted vs. original signals and MSE for ECG signal from MIT-BIH arrhythmia database for Sample 207 and lead MLII, d=4, T=5.0, and $M_p=107, 995$.  Modeling phase is $\in[0,50]$ secs. for M=18,000 training data. New data predicted $\in[50,300]$ secs. is $M_p-M$=89,995.  }
\end{figure}
\newpage

\newpage
\begin{figure}[ht]
 \centering
 \subfigure[Predicted vs. original signals.]{
  \includegraphics[width =1.0\linewidth]{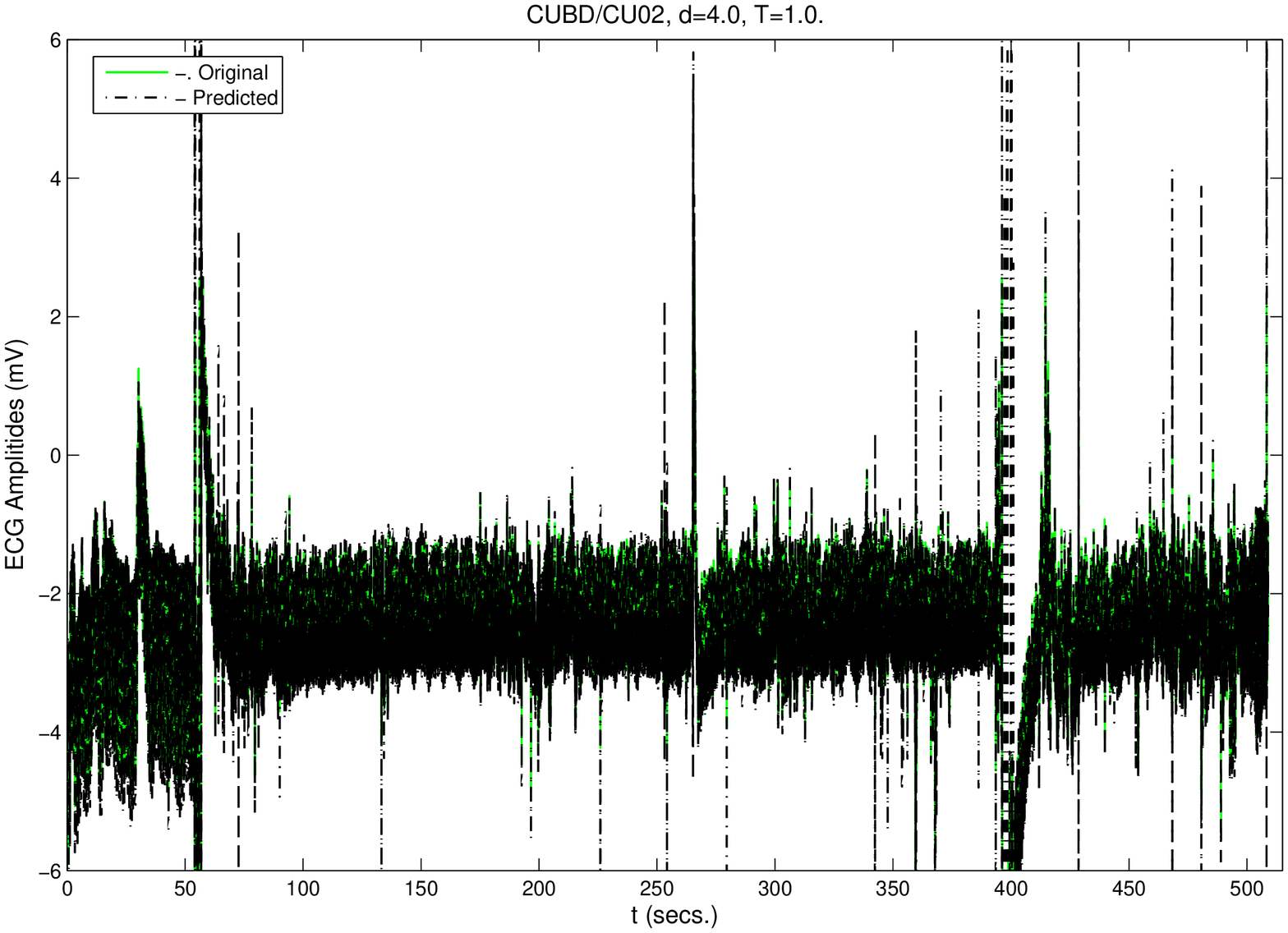}
   \label{fig:subfig1}
   }
 \subfigure[MSE]{
  \includegraphics[width =1.0\linewidth]{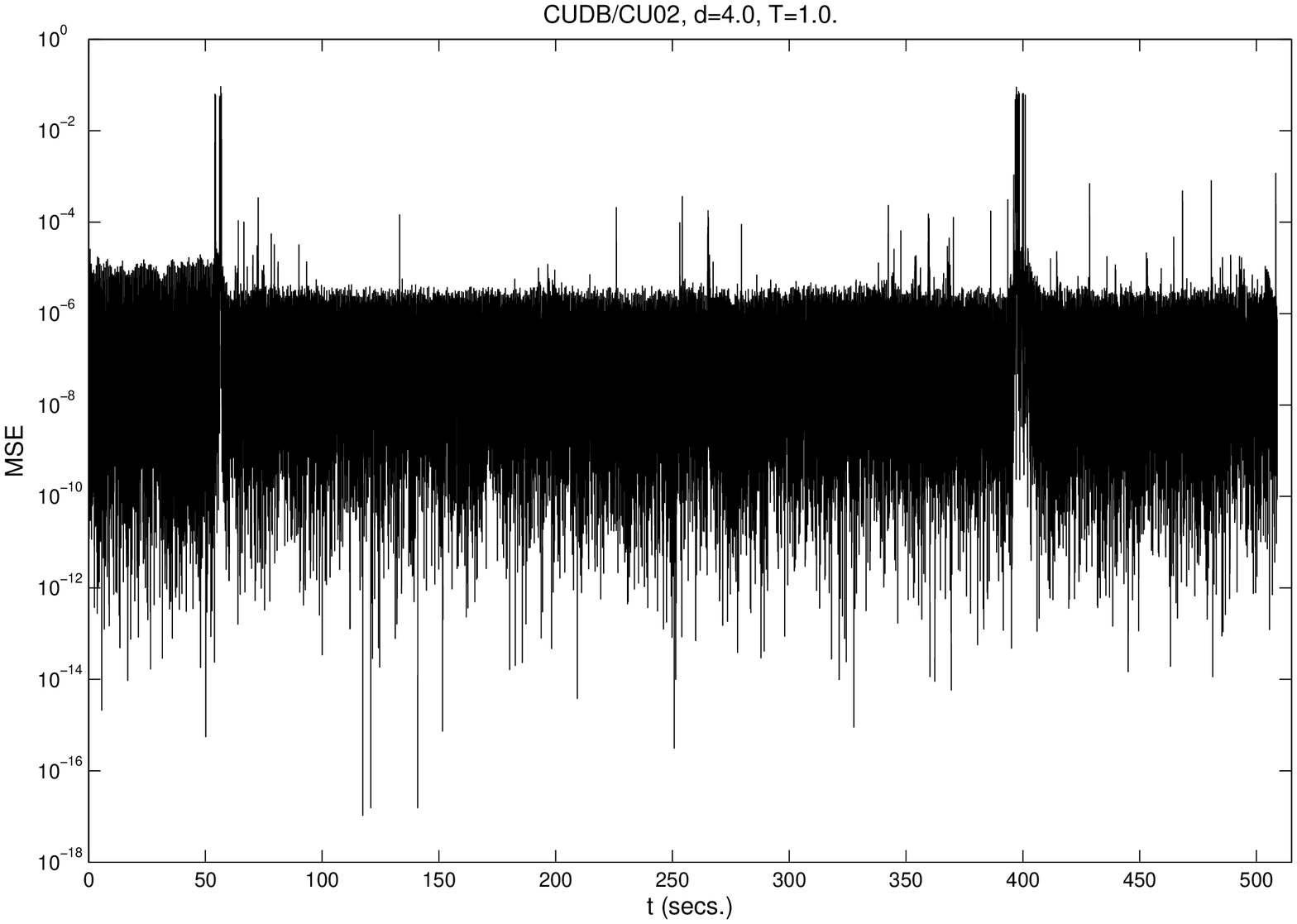}
   \label{fig:subfig1}
   }

\label{fig:subfigureExample}
 \caption[Optional caption for list of figures]{%
   Predicted vs. original signals and MSE for ECG signal from Creighton University ventricular tachyarrhythmia database Sample CUDB/CU02, d=4, T=1.0, and $M_p=127,228$.  Modeling phase is $\in[0,120]$ secs. for M=30,000 training data. New data predicted $\in[120,515]$ secs. is $M_p-M$=97,228.  }
\end{figure}

\begin{figure}[ht]
 \centering
 \subfigure[Predicted vs. original signals.]{
  \includegraphics[width =1.0\linewidth]{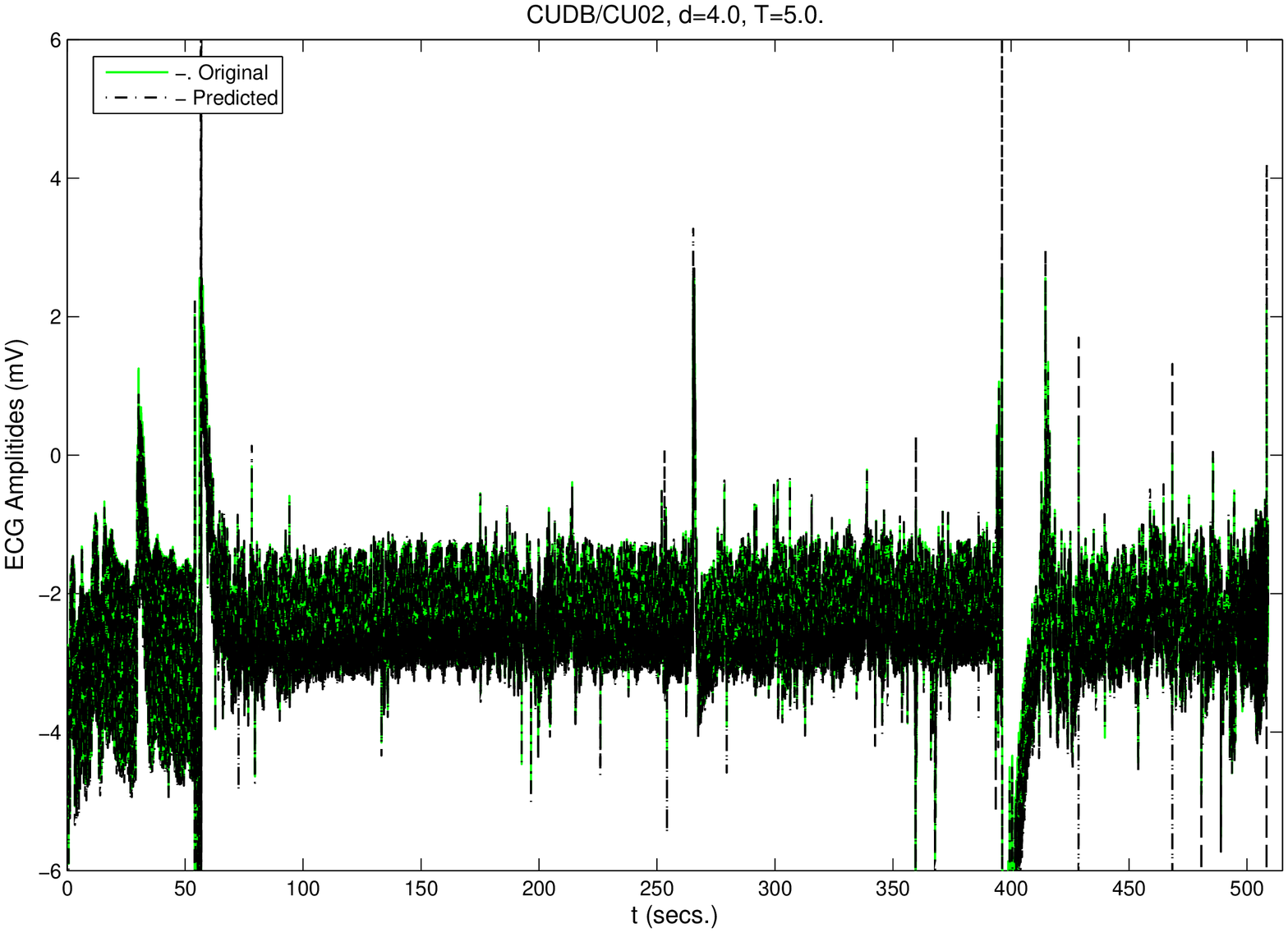}
   \label{fig:subfig1}
   }
 \subfigure[MSE]{
  \includegraphics[width =1.0\linewidth]{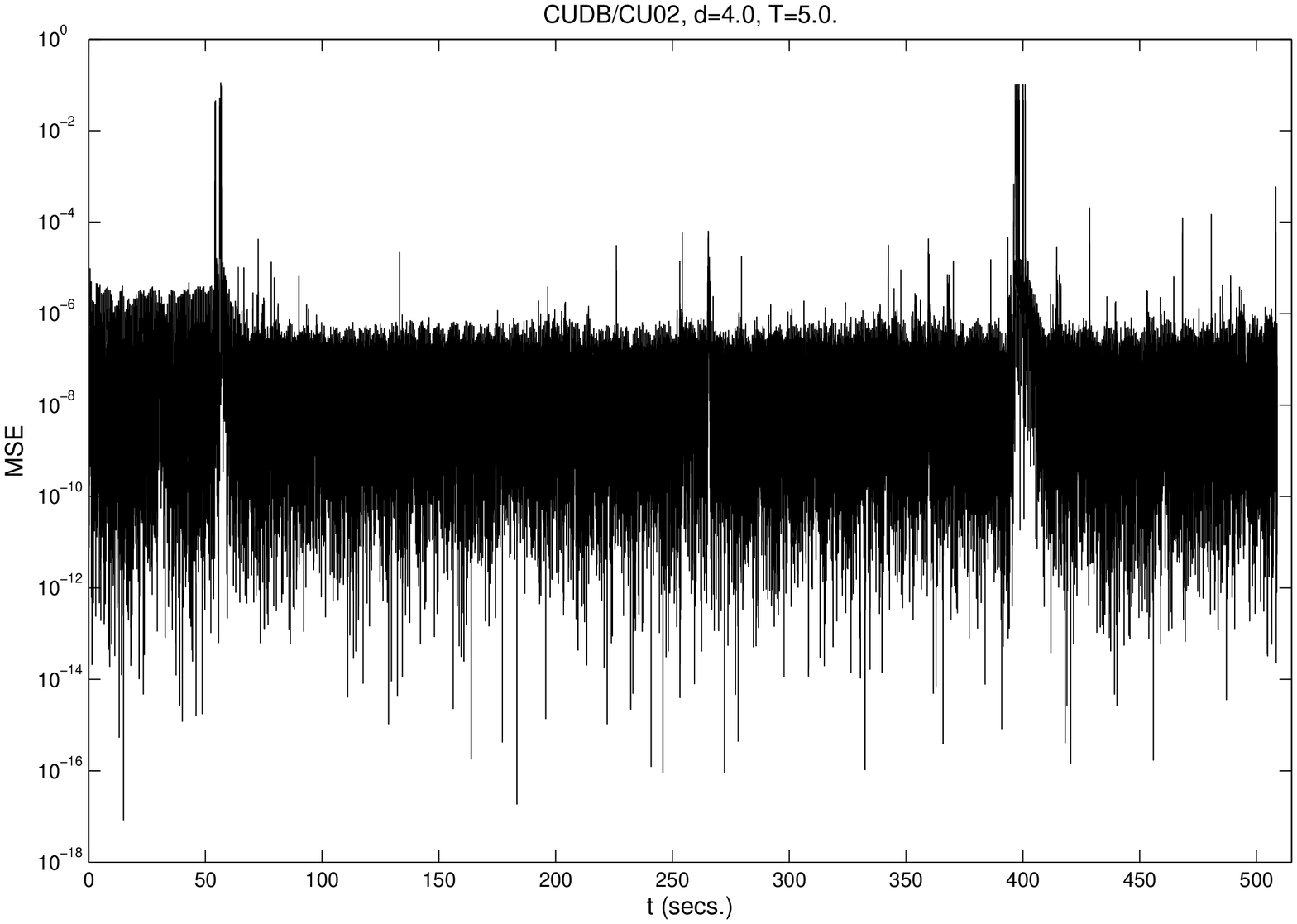}
   \label{fig:subfig1}
   }

\label{fig:subfigureExample}
 \caption[Optional caption for list of figures]{%
 Predicted vs. original signals and MSE for ECG signal from Creighton University ventricular tachyarrhythmia database Sample CUDB/CU02, d=4, T=5.0, and $M_p=127,227$.  Modeling phase is $\in[0,120]$ secs. for M=30,000 training data. New data predicted $\in[120,515]$ secs. is $M_p-M$=97,227.  }
\end{figure}

\end{document}